\documentclass[a4paper,fleqn,usenatbib]{mnras}

\usepackage{newtxtext,newtxmath}

\usepackage[T1]{fontenc}
\usepackage{ae,aecompl}

\usepackage{graphicx}	
\usepackage{amsmath}	
\usepackage{amssymb}	

\usepackage{subfig}
\usepackage{xcolor}

\title[Extragalactic proper motions]{Cosmology with extragalactic proper motions: harmonic formalism, estimators, and forecasts}

\author[Alex Hall]{
Alex Hall$^{1}$\thanks{E-mail: ahall@roe.ac.uk}
\\
$^{1}$Institute for Astronomy, University of Edinburgh, Royal Observatory, Blackford Hill, Edinburgh, EH9 3HJ, UK
}

\date{Accepted XXX. Received YYY; in original form ZZZ}

\pubyear{2018}

\begin{document}
\label{firstpage}
\pagerange{\pageref{firstpage}--\pageref{lastpage}}
\maketitle

\begin{abstract}
We conduct a thorough study into the feasibility of measuring large-scale correlated proper motions of galaxies with astrometric surveys. We introduce a harmonic formalism for analysing proper motions and their correlation functions on the sphere based on spin-weighted spherical harmonics, and study the statistics of the transverse velocity field induced by large-scale structure. We use a likelihood formalism to derive optimal estimators for the secular parallax due to the Solar System's motion relative to distant objects, and compute the variance and bias due to peculiar velocities and relativistic aberration. We use a simulated catalogue of galaxy proper motions with radial distributions and noise properties similar to those expected from Gaia to forecast the detectability of the proper motion dipole, whose amplitude may be considered a proxy for the Hubble constant. We find cosmic variance to be the limiting source of noise for this measurement, forecasting a detectability of $1$-$2\sigma$ on a single component of the local velocity, increasing to $2$-$4\sigma$ (equivalent to a 25\%-50\% measurement of the Hubble constant) if the CMB dipole is included as prior information. We conduct a thorough study into the radial dependence of the signal-to-noise, finding that most of the information comes from galaxies closer than a few hundred Mpc. We forecast that the amplitude of peculiar transverse velocities can potentially be measured with 10$\sigma$ significance; such a measurement would offer a unique probe of cosmic flows and a valuable test of the cosmological model.
\end{abstract}




\section{Introduction}
\label{sec:intro}

The launch of the ESA Gaia\footnote{\url{http://sci.esa.int/gaia/}} satellite and its subsequent data releases are expected to revolutionize the field of astrometry, producing the largest catalogue of precise positions and proper motions to date. Upcoming very-long baseline interferometry measurements with the ngVLA\footnote{\url{http://ngvla.nrao.edu/}} will also have unprecedented astrometric precision at radio frequencies. These experiments should both have end-of-mission proper motion accuracy of order $10 \, \mu\mathrm{as} \, \mathrm{yr}^{-1}$ for the brightest objects.

With the dramatic increase in data volume and quality it is timely to ask what precision astrometry can say about cosmology. Although not designed for this purpose, Gaia will measure the proper motions of some $10^6$ galaxies~\citep{2014A&A...568A.124D, 2015A&A...576A..74D}, preferentially selecting objects which look most `point-source-like', i.e. ellipticals with large bulge-to-disk components. Additionally, a large population of quasars will be observed to pin down the celestial reference frame (see, e.g.~\citealt{2018A&A...616A..14G}).

There are several potential uses of such a dataset for cosmology (see~\citealt{2018arXiv180706670D} for a review). Firstly, since the Solar System moves relative to distant objects, there is a `secular parallax' (SP) proper motion in the opposite direction to our local velocity. This proper motion has an amplitude of roughly $\frac{80}{r / 1 \, \mathrm{Mpc}} \, \mu\mathrm{as} \, \mathrm{yr}^{-1}$, where $r$ is the comoving distance to the object, and has a dipolar dependence on angle, anti-aligned with the velocity vector of the Solar System with respect to the CMB rest frame\footnote{We will often express quantities in the CMB rest frame, the frame in which the CMB dipole vanishes~\citep{2014A&A...571A..27P}. We assume that the matter rest frame (the frame in which the dipole anisotropy of peculiar velocities vanishes) and the CMB rest frame coincide, i.e. we do not consider the `tilted universe' scenario of~\citet{1991PhRvD..44.3737T}. We also neglect any intrinsic dipole anisotropy in the CMB.}. With spectroscopic redshifts as a proxy for distance, the SP proper motion may be used to infer the Hubble constant, $H_0$, if the SP velocity is fixed by the CMB dipole. \citet{2016A&A...589A..71B} forecast that Gaia can potentially make a $\sim30\%$ measurement of $H_0$ with this method, with further improvements possible if more galaxies can be detected. A competitive measurement of $H_0$ could shed light on recent tensions between CMB and classical distance-ladder measurements (see~\citealt{2017NatAs...1E.121F} for a review). At greater distances, quasars may be used to probe the distance-redshift relation to constrain dark energy~\citep{2009MNRAS.397.1739D}.

Complicating the measurement of SP is the typically larger signal of the `secular aberration drift' (SAD) proper motion due to the time-varying relativistic aberration from the acceleration of the Solar System towards the galactic centre. This has a magnitude of roughly $4  \, \mu\mathrm{as} \, \mathrm{yr}^{-1}$, with a dipolar angular dependence directed towards the galactic centre, and has been measured in quasars by~\citet{2011A&A...529A..91T}. This signal is independent of distance, which in principle allows it to be distinguished from the SP proper motion; alternatively it can be measured from high-redshift quasars where SP is negligible and then subtracted.

Going beyond local effects, galaxies and quasars have intrinsic peculiar velocities caused by large-scale structure (LSS). In linear perturbation theory this gives rise to an r.m.s. proper motion of roughly $\frac{90}{r / 1 \, \mathrm{Mpc}} \, \mu\mathrm{as} \, \mathrm{yr}^{-1}$. This is roughly the same size as the SP proper motion and has the same distance dependence, and represents an important source of bias and variance in attempts to measure SP (unsurprisingly, as we are only sensitive to the relative motion between the Solar System and extragalactic objects). However, this signal has a distinct correlation structure, with quadrupolar, octupolar, and higher-order angular structure due to correlations in the peculiar velocity field. This angular dependence in principle allows LSS transverse velocities to be partly separated from the SP effect, which would provide a valuable probe of large-scale motions, free from the Malmquist biases that affect radial velocity surveys~\citep{2012ApJ...755...58N}. Cosmic velocity fields are sensitive to large-scale inhomogeneities in the dark matter density field which can be modelled accurately in linear perturbation theory; such a measurement could provide a valuable probe of late-time physics such as dark energy. Predictions for the sensitivity of Gaia to these motions range from $1$-$2\sigma$~\citep{2012ApJ...755...58N} to $10\sigma$~\citep{2018arXiv180706670D}, depending on survey assumptions and measurement techniques.

In this work, we extend previous studies into the feasibility of detecting large-scale correlated proper motions, paying particular attention to the `cosmic variance' imparted by LSS. Proper motions from SP and LSS are strongest in nearby objects, which is also the regime where correlations between the relevant velocities are expected to be strongest. Despite this, previous studies have adopted only a simplistic approach to including cosmic variance in their forecasts, and have typically neglected correlations between transverse velocities. We consistently account for these correlations, and identify a previously-neglected source of bias in the measurement of the proper motion dipole coming from correlations between the Solar System's motion and those of nearby galaxies, closely related to the `bulk flow' phenomenon observed in radial velocity surveys.

We adopt a statistical approach to measuring large-scale correlated proper motions based on a likelihood function, and use it to derive optimal `stacking' estimators for SP proper motion and its LSS counterpart. We present several mitigation strategies to reduce the bias and variance of these estimators. To do this, we introduce a CMB-style formalism for measuring correlated proper motions across the full sky, which serves as an alternative to the widely-used Vector Spherical Harmonics (VSH;~\citealt{2012A&A...547A..59M}). This formalism uses the spin-weighted spherical harmonic decomposition of the proper motion field, which allows correlation functions and power spectra to be easily constructed and analysed (extending recent work by~\citealt{2018ApJ...864...37D}), and ensures that all the information in the vector field is used. This formalism is easier to use than VSH when correlating proper motions, and we expect it to be useful beyond the applications presented here.

We also conduct a detailed study into the statistics of transverse velocities, focussing on their redshift-dependence, angular structure, and sensitivity to non-linearity in dark matter inhomogeneities. Our likelihood formalism allows us to consistently propagate the variance from LSS velocities through to estimators of $H_0$ and the amplitude of the peculiar velocity field. We present forecasts for these quantities with a Gaia-like astrometric survey, and thoroughly investigate the radial dependence of the signal-to-noise.

This paper is structured as follows. In Section~\ref{sec:harmonics} we introduce the spin-weighted spherical harmonics as a means of analysing transverse vectors on the full sky, and describe how correlation functions and power spectra may be constructed and compared with theoretical models such as linear perturbation theory. In Section~\ref{sec:like} we construct a likelihood function for proper motions, and use it to derive optimal estimators for large-scale correlated signals. In Section~\ref{sec:gums} we describe our Gaia-like simulated galaxy catalogue, and in Section~\ref{sec:results} we present forecasts for large-scale proper motion dipoles and $H_0$. In Section~\ref{sec:LSS} we extend our formalism to the measurement of transverse velocities induced by large-scale structure, and we conclude in Section~\ref{sec:concs}. In a series of appendices we present further technical details of our harmonic formalism.

We set the speed of light $c=1$ throughout this work. All results were computed with the best-fitting flat $\Lambda$CDM cosmological parameters (TT, EE, TE + lowP + lensing + ext) quoted in~\citet{2016A&A...594A..13P}, namely $(\Omega_bh^2, \Omega_ch^2, h, A_s, n_s) = (0.0223, 0.1188, 0.6774, 2.142 \times 10^{-9}, 0.9667)$.

\section{Power spectra and correlation functions of proper motion}
\label{sec:harmonics}

In this section we present the harmonic decomposition of a transverse velocity or proper motion field on the sphere, and construct the correlation functions and power spectra which form the basis of our likelihood formalism. Much of the material here will be familiar from CMB polarization studies, and further technical details are provided in Appendix~\ref{app:formalism}. Some of the formulae here are also presented in Appendix B of~\citet{2014PhRvD..90f3518H}, albeit in a different context.

\subsection{Rotational properties of the transverse velocity}
\label{subsec:vrot}

When dealing with vector or tensor quantities on the sphere, it is usually advantageous to work with objects which do not depend on a coordinate system, since any physically meaningful quantity cannot depend on the adopted coordinate system. Independence from the coordinate system is most easily achieved by working with objects which transform straightforwardly under a change of basis.

In the case of transverse velocity (or proper motion, related to transverse velocity by a factor of $1/r$), we commonly work with a locally orthonormal pair of basis vectors $(\hat{\mathbf{x}},\hat{\mathbf{y}})$, orthogonal to the line of sight. Denoting by $\hat{\mathbf{n}}$ the outward radial unit vector corresponding to the line of sight, the vectors $(\hat{\mathbf{x}},\hat{\mathbf{y}}, -\hat{\mathbf{n}})$ form a right-handed set and the components of the transverse velocity in this system are $(V_x, V_y)$. By considering the standard two-dimensional rotation matrix, one sees that under a right-handed rotation of the coordinate basis about $-\hat{\mathbf{n}}$ (or left-handed about $\hat{\mathbf{n}}$) by an angle $\gamma$, the complex transverse velocity $V_x + iV_y$ transforms as
\begin{equation}
  V_x + iV_y \rightarrow (V_x + iV_y)e^{i\gamma}.
  \label{eq:vperp_rotate}
\end{equation}
Now, we say a quantity ${}_s \eta$ has spin $s$ if, under the transformation $\hat{\mathbf{x}} + i\hat{\mathbf{y}} \rightarrow (\hat{\mathbf{x}} + i\hat{\mathbf{y}})e^{i\gamma}$, we have ${}_s \eta \rightarrow e^{is\gamma}{}_s\eta$. The complex transverse velocity $V_x \pm iV_y$ thus has spin $\pm 1$. In this work we use the standard orthonormal spherical polar basis vectors $(\hat{\boldsymbol{\theta}},\hat{\boldsymbol{\phi}}, \hat{\mathbf{n}})$, which form a right-handed set, and define the complex components of the transverse velocity as $V_{\pm} \equiv V_{\theta} \pm i V_{\phi}$. The components in this basis are related to those in the celestial reference frame (described by declination $\delta$ and $\cos{\delta}$-corrected right-ascension $\alpha^*$) by $(V_{\theta}, V_{\phi}) = (-V_{\delta}, V_{\alpha^*})$. We prefer to work with the quantities $V_{\pm}$ over $(V_{\theta}, V_{\phi})$ because their simple transformation behaviour under rotations makes constructing correlation functions considerably easier, as we shall see.

Being a spin $\pm 1$ field on the sphere, $V_{\pm}$ may naturally be expanded in spin $\pm 1$ spherical harmonics (a set of orthonormal basis functions on the sphere possessing the correct rotational properties, see Appendix~\ref{app:formalism}) as
\begin{equation}
  V_{\pm}(\hat{\mathbf{n}}) = \sum_{lm} (\mp \epsilon_{lm} + i \beta_{lm}) {}_{\pm 1}Y_{lm}(\hat{\mathbf{n}}),
  \label{eq:vexp}
\end{equation}
where the sum over $m$ ranges between $-l$ and $l$ and we have $l \ge 1$. As discussed in Appendix~\ref{app:formalism}, under a parity transformation the quantities $\epsilon_{lm}$ transform as $\epsilon_{lm} \rightarrow (-1)^l \epsilon_{lm}$ and are hence said to have electric parity, whereas the $\beta_{lm}$ transform as $\beta_{lm} \rightarrow (-1)^{l+1} \beta_{lm}$ and hence have magnetic parity. It is straightforward to show that these objects are related to the solenoidal ($s_{lm}$) and toroidal ($t_{lm}$) coefficients of a Vector Spherical Harmonic (VSH) expansion~\citep{2012A&A...547A..59M} by $(s_{lm}, t_{lm}) = (\epsilon_{lm}, \beta_{lm})$.

\subsection{Proper motion dipoles}
\label{subsec:dipoles}

In this work we are primarily concerned with measuring proper motions which are correlated over large angular scales, such as the dipolar secular parallax. In this section we show how to map a proper motion or transverse velocity dipole to the multipole coefficients $\epsilon_{lm}$ and $\beta_{lm}$.

In the secular parallax scenario, the Solar System moves with some velocity $\mathbf{V}$ relative to distant objects, which are assumed fixed. Taking the $z$-axis of a spherical coordinate system along $\mathbf{V}$, the transverse velocity of distant galaxies or quasars relative to the Solar System barycentre is $\mathbf{V}_\perp = \lvert \mathbf{V} \rvert \sin \theta' \hat{\boldsymbol{\theta}}'$, where a prime denotes quantities in the frame with $\hat{\mathbf{z}} = \hat{\mathbf{V}}$. In this coordinate system then we have $(V_\theta',V_\phi') = (\lvert \mathbf{V} \rvert \sin \theta', 0)$. Using the explicit expressions for the spin-weighted spherical harmonics in Equation~\eqref{appeq:1Y0} and the definition of the complex transverse velocity we thus have
\begin{equation}
  V_{\pm}'(\hat{\mathbf{n}}) = \pm \lvert \mathbf{V} \rvert \sqrt{\frac{8\pi}{3}}{}_{\pm1}Y_{10}(\hat{\mathbf{n}}).
\end{equation}
We can now use Equation~\eqref{eq:vexp} to read off the multipoles in this frame, finding $\epsilon_{lm}' = -\lvert \mathbf{V} \rvert \sqrt{8\pi/3}\delta_{l1}\delta_{m0}$ and $\beta_{lm}' = 0$. Thus, in this frame, the secular parallax is purely E-mode and dipolar. This example illustrates a generic feature of the E-mode; the direction in which the amplitude of the transverse velocity is maximally changing is parallel or anti-parallel to the local direction of the transverse velocity. If instead we rotated the transverse velocity vector at each point by $90^\circ$, we would have a pure B-mode pattern characteristic of global rotation of distant objects about the $z$-axis. In this case the amplitude is maximally changing $90^\circ$ or $270^\circ$ to the local direction, characteristic of a pure B-mode.

To find the multipole coefficients in a general frame, we use the rotation law in Equation~\eqref{appeq:rotelm}, $\epsilon_{lm} = \sum_{m'}D^l_{mm'}(\alpha,\beta,\gamma) \epsilon_{lm'}'$ where $(\alpha, \beta, \gamma)$ are the Euler angles which rotate from the primed frame to the unprimed (general) frame, and the $D^l_{mm'}$ are the Wigner $D$ matrix elements (see Appendix~\ref{app:formalism}). The velocity vector in a general frame has direction $\hat{\mathbf{V}} = (\sin \beta \cos \alpha, \sin \beta \sin \alpha, \cos \beta)$, so using the relation of the Wigner $D$ matrix elements to the spin-weighted spherical harmonics in Equation~\eqref{appeq:D2SH} we have, in a general frame
\begin{align}
  \epsilon_{lm} &= -\frac{8\pi}{3}\frac{\lvert \mathbf{V} \rvert}{\sqrt{2}} Y_{1m}^*(\hat{\mathbf{V}})\delta_{l1}, \nonumber \label{eq:elmdip}\\
  \beta_{lm} &= 0.
\end{align}
Note that the rotation has preserved the pure E-mode nature of the signal, a generic feature of the E/B decomposition.

To gain further intuition into the multipole coefficients, consider inserting the explicit forms of the spherical harmonics into Equation~\eqref{eq:elmdip}. This yields
\begin{align}
  \epsilon_{1-1} &= -\sqrt{\frac{8\pi}{3}}\frac{V_x + iV_y}{\sqrt{2}}, \nonumber \\
  \epsilon_{10} &= -\sqrt{\frac{8\pi}{3}} V_z \nonumber, \\
  \epsilon_{11} & = -\sqrt{\frac{8\pi}{3}}\frac{-V_x + iV_y}{\sqrt{2}}.
\end{align}
Thus, the multipole coefficients are just the components of the spatially fixed three-dimensional velocity field in a helicity basis. Writing $\boldsymbol{\epsilon} = (\epsilon_{1-1},\epsilon_{10},\epsilon_{11})^\intercal$, this may be compactly written as $\boldsymbol{\epsilon} = -\sqrt{8\pi/3}\mathbfss{B}{\mathbf{V}}$, with $\mathbfss{B}$ a unitary matrix.
  
\subsection{Correlation functions}
\label{subsec:corrfuncs}

Correlating vectors on the sphere is slightly more subtle than correlating scalars, since one must ensure the final result is independent of the orientation of the $(\hat{\mathbf{x}},\hat{\mathbf{y}})$ basis, which itself varies over the sky. Examples of basis-independent transverse-velocity correlation functions were recently presented by~\citet{2018ApJ...864...37D}, where it was pointed out that several choices of correlation function may be made but any single choice is not sufficient to capture all the information in the vector field, which has two degrees of freedom at each point.

When correlating a vector field at two point on the sky, the natural basis in which to express the components is provided by the geodesic connecting the two points. Let $\gamma_1$ be the angle required to rotate the local $(\hat{\boldsymbol{\theta}},\hat{\boldsymbol{\phi}})$ basis at $\hat{\mathbf{n}}_1$ in a right-handed sense about $\hat{\mathbf{n}}_1$ such that the $\hat{\boldsymbol{\theta}}$ vector is aligned with the geodesic connecting $\hat{\mathbf{n}}_1$ and $\hat{\mathbf{n}}_2$ Let $\gamma_2$ be the corresponding angle at $\hat{\mathbf{n}}_2$. On the flat sky we have $\gamma_1 = \gamma_2$. Let $\cos \beta_{12} = \hat{\mathbf{n}}_1 \cdot \hat{\mathbf{n}}_2$. The complex transverse velocity in this rotated basis, $\bar{V}_{\pm}$ may be found using Equation~\eqref{eq:vperp_rotate}. It is related to the complex transverse velocity in the global basis by
%
%
%
%
\begin{equation}
  \bar{V}_{\pm}(\hat{\mathbf{n}}_1) = V_{\pm}(\hat{\mathbf{n}}_1)e^{\mp i \gamma_1},
\end{equation}
and likewise at $\hat{\mathbf{n}}_2$. The angles $\{ \gamma_1, \beta_{12}, -\gamma_2\}$ form a set of Euler angles which rotate the basis at $\hat{\mathbf{n}}_1$ into that at $\hat{\mathbf{n}}_2$.

Given two points $(r_1, \hat{\mathbf{n}}_1)$ and $(r_2, \hat{\mathbf{n}}_2)$ in three-dimensional space\footnote{Throughout this work, $r$ refers to comoving distance, which is equal to the `proper motion distance' which links transverse velocity to proper motion in a spatially flat universe~\citep{1999astro.ph..5116H}.}, there are two correlation functions we can form with $\bar{V}_{\pm}(\hat{\mathbf{n}}_1, r_1)$ and $\bar{V}_{\pm}(\hat{\mathbf{n}}_2, r_2)$. Assuming that the transverse velocity field has vanishing mean (which is true for cosmic velocities sourced by large-scale structure), these are given by
\begin{align}
  \xi_\pm(\beta_{12}, r_1, r_2) &\equiv  \langle \bar{V}_+(\hat{\mathbf{n}}_1, r_1) \bar{V}_{\pm}^*(\hat{\mathbf{n}}_2, r_2) \rangle \nonumber \\
  &= e^{-i\gamma_1}e^{\pm i\gamma_2} \langle V_+(\hat{\mathbf{n}}_1, r_1) V_{\pm}^*(\hat{\mathbf{n}}_2, r_2) \rangle,
  \label{eq:xipm}
\end{align}
where the angle brackets denote an expectation value over realizations of the transverse velocity field. As we shall see, these correlation functions are real-valued.

We can relate the correlation functions $\xi_\pm$ to the power spectra of the multipole coefficients by plugging Equation~\eqref{eq:vexp} into Equation~\eqref{eq:xipm}. First we define the power spectra
\begin{align}
  \langle \epsilon_{lm}(r_1) \epsilon^*_{l'm'}(r_2) \rangle &= \zeta^E_l(r_1,r_2) \delta_{ll'}\delta_{mm'}, \nonumber \\
  \langle \beta_{lm}(r_1) \beta^*_{l'm'}(r_2) \rangle &= \zeta^B_l(r_1,r_2) \delta_{ll'}\delta_{mm'}, \nonumber \\
  \langle \epsilon_{lm}(r_1) \beta^*_{l'm'}(r_2) \rangle &= 0.
\end{align}
where we have assumed statistical isotropy (which imposes the Kronecker deltas and the $m$-independence of the power spectra) and parity non-violation (which ensures that the EB correlation vanishes). Note that these power spectra are real-valued. Inserting these definitions into Equation~\eqref{eq:xipm} give
\begin{align}
  \xi_\pm(\beta_{12}, r_1, r_2) &= e^{-i\gamma_1}e^{\pm i\gamma_2} \sum_l \left(\pm \zeta^E_l(r_1,r_2) + \zeta^B_l(r_1,r_2) \right) \nonumber \\
  & \, \times \sum_m {}_1 Y_{lm}(\hat{\mathbf{n}}_1) {}_{\pm 1}Y_{lm}^*(\hat{\mathbf{n}}_2).
  \label{eq:xipmlms}
\end{align}
Using Equation~\eqref{appeq:YD} to relate the spin-weighted spherical harmonics to the Wigner $D$ matrix elements~\citep{Varsh} we find
\begin{equation}
  \xi_\pm(\beta_{12}, r_1, r_2) = \sum_l \frac{2l+1}{4\pi} \left [ \pm \zeta^E_l(r_1,r_2) + \zeta_l^B(r_1,r_2) \right] d^l_{1\pm1}(\beta_{12}),
  \label{eq:xipmzeta}
\end{equation}
where $d^l_{1\pm1}(\beta_{12})$ are elements of the reduced Wigner matrices, and are the generalization of the Legendre polynomials to fields with non-zero spin. These functions are real-valued and obey the relations $d^l_{1-1} = d^l_{-1 1}$ and $d^l_{11} = d^l_{-1-1}$, and may be computed using the recursion relation~\citep{Varsh}
\begin{equation}
  d^l_{1 \pm 1} = \frac{l(2l-1)}{l^2-1}\left[\left(\cos \beta \mp \frac{1}{l(l-1)}\right)d^{l-1}_{1 \pm1} - \frac{l(l-2)}{(l-1)(2l-1)}d^{l-2}_{1 \pm 1}\right],
\end{equation}
for $l\ge3$, with the boundary conditions
\begin{align}
  d^1_{1 \pm 1} &= \frac{1}{2}(1 \pm \cos \beta), \nonumber \\
  d^2_{1 \pm 1} &= \frac{1}{2}(\pm 2\cos^2 \beta + \cos \beta \mp 1).
\end{align}
The Wigner $d$ elements obey the relations $d^l_{11}(0) = 1$ and $d^l_{1-1}(0) = 0$, which means that the quantity $(2l+1)(\zeta^E_l + \zeta^B_l)/4\pi$ is the contribution per $l$ to the variance at a point, and the quantity $l(l+1)(\zeta^E_l + \zeta^B_l)/2\pi$ is roughly the contribution per $\log l$ to the variance at a point.

Since the $d$ functions and the power spectra are real, the $\xi_\pm$ are real and hence completely describe the correlation structure of the transverse velocity field. If in addition the transverse velocity is Gaussian distributed with zero mean, these correlation functions completely describe the statistics of the vector field.

In Appendix~\ref{app:darling} we compare the $\xi_\pm$ correlation functions with those introduced by~\citep{2018ApJ...864...37D}, finding that the two sets are linear combinations of each other.

\subsection{The transverse velocity of large-scale structure}
\label{subsec:LSSspec}

Large-scale gravitational potentials induce departures from the Hubble flow in the motion of galaxies and dark matter haloes. Since the initial conditions for structure formation appear Gaussian to a close approximation, on large scales the velocity field of large-scale structure also appears Gaussian due to the evolution being close to linear. On small scales the effects of non-linear evolution impart non-Gaussianity into the peculiar velocity field $\mathbf{v}$, but the effects are quite weak for wavenumbers $k \lesssim 0.5 \, h \, \mathrm{Mpc}^{-1} $ at $z=0$~\citep{2013PhRvD..88j3510Z}. Unlike the matter overdensity $\delta$, the peculiar velocity is not required to be greater than $-1$, and so non-Gaussianity is not a fundamental requirement when fluctuations in $\mathbf{v}$ become large~\citep{1991MNRAS.248....1C}.

Gaussian fields are completely described by their mean and covariance, and since the former is zero by definition for peculiar velocity, the power spectrum or correlation function of $\mathbf{v}$ is of central interest in studying cosmic flows. In particular, the proper motion of LSS can act as a contaminant for measurements of the Hubble constant from secular parallax, and so quantifying its statistics is a key goal of this paper. The LSS proper motion is also potentially measurable in astrometric surveys~\citep{2012ApJ...755...58N}, offering the possibility to constrain cosmological parameters such as the dark energy equation of state from real-time cosmology. In this section, we compute the power spectrum and correlation functions of peculiar velocities, focussing on the low redshifts relevant for proper motion surveys. Some of this material has been presented in~\citep{2014PhRvD..90f3518H} in the context of polarized emission from the kinetic Sunyaev-Zeldovich effect, but we repeat it here for completeness.

On the large scales relevant for this work, the peculiar velocity field is, to a good approximation, curl-free~\citep{2009MNRAS.393..297P}. As described in Appendix~\ref{app:formalism}, this implies that the transverse velocity is a pure E-mode, described by a single power spectrum $\zeta^E_l(r_1,r_2)$ for galaxies located at radii $r_1$ and $r_2$. In Fourier space we have $\mathbf{v}(\mathbf{k}) = -i\hat{\mathbf{k}}v(\mathbf{k})$ where $v(\mathbf{k})$ is the velocity potential. The real-space peculiar velocity is then
\begin{equation}
  \mathbf{v}(\mathbf{r}) = \nabla_{\mathbf{r}} \int \frac{\mathrm{d}^3 \mathbf{k}}{(2\pi)^3} \, \frac{-v(\mathbf{k})}{k} e^{i\mathbf{k} \cdot \mathbf{r}},
\end{equation}
where $\nabla_{\mathbf{r}}$ is the three-dimensional spatial gradient. Projecting onto the sphere with $\mathbf{r} = r\hat{\mathbf{n}}$ and using that $\nabla^\perp_{\mathbf{r}} = (1/r)\nabla$ where $\nabla$ is the angular covariant derivative, we find $v^a = \nabla^a \Omega$ with
\begin{equation}
  \Omega(\hat{\mathbf{n}},r) = -\int \frac{\mathrm{d}^3 \mathbf{k}}{(2\pi)^3} \, \frac{v(\mathbf{k},r)}{kr} e^{ikr\hat{\mathbf{k}} \cdot \hat{\mathbf{n}}}.
\end{equation}
Expanding the exponential in spherical harmonics and using the multipole expansion in Equation~\eqref{appeq:omegalm} allows us to read off the E-mode coefficients for the transverse velocity as
\begin{equation}
  \epsilon_{lm}(r) = -4\pi i^l\sqrt{l(l+1)} \int  \frac{\mathrm{d}^3 \mathbf{k}}{(2\pi)^3} \, v(\mathbf{k},r) \frac{j_l(kr)}{kr} Y^*_{lm}(\hat{\mathbf{k}}),
\end{equation}
where $j_l$ is a spherical Bessel function.

We now use the Newtonian linear continuity equation to write $v(\mathbf{k},r) = -[\mathcal{H(\eta)}/k]f(\eta)\delta(\mathbf{k},r)$ where $\mathcal{H}(\eta)$ is the comoving Hubble parameter at conformal time $\eta = \eta_0 - r$, with $\eta_0$ the current conformal time, and $f(\eta)$ is the growth rate of linear density fluctuations. We neglect any velocity bias, which has been shown to be very small on the relevant spatial scales here~\citep{2018ApJ...861...58C}. Using the statistical isotropy and homogeneity of the density field allows us to compute the E-mode angular power spectrum as
\begin{align}
  \zeta^E_l(r_1,r_2) &= 4\pi l(l+1)\mathcal{H}(\eta_1)\mathcal{H}(\eta_2)f(\eta_1)f(\eta_2) \nonumber \\
  & \, \times \int \frac{k^2\mathrm{d}k}{2\pi^2} \, \frac{P(k;\eta_1,\eta_2)}{k^2} \frac{j_l(kr_1)}{kr_1}\frac{j_l(kr_2)}{kr_2},
  \label{eq:zetaE}
\end{align}
where $P(k; r_1,r_2)$ is the unequal-time matter power spectrum. The correlation functions $\xi_\pm$ may now be found using Equation~\eqref{eq:xipmzeta} with $\zeta^B_l = 0$.

\begin{figure}
  \includegraphics[width=\columnwidth]{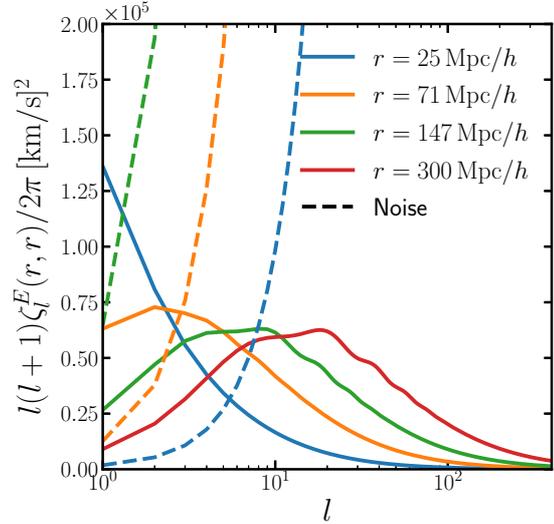}
  \caption{The angular power spectrum of LSS transverse velocities in linear theory, at (highest to lowest at $l=1$) $r = 25 \, \mathrm{Mpc}/h$ (blue solid), $r = 71 \, \mathrm{Mpc}/h$ (orange solid), $r = 147 \, \mathrm{Mpc}/h$ (green solid), and $r = 300 \, \mathrm{Mpc}/h$ (red solid). We also plot the instrumental noise power for  $r = 25 \, \mathrm{Mpc}/h$ (lowest, blue dashed), $r = 71 \, \mathrm{Mpc}/h$ (middle, orange dashed), and $r = 147 \, \mathrm{Mpc}/h$ (highest, green dashed). The correlation function at a given angular separation is found by summing over multipoles of these power spectra weighted by the Wigner $d$ functions, as in Equation~\eqref{eq:xipmzeta}. Note that BAOs become increasingly visible at greater distances. Structure at greater distance has power on smaller angular scales due to the roughly constant coherence length of the three-dimensional velocity field.}
  \label{fig:vperp_ells_xiE}
\end{figure}

In Figure~\ref{fig:vperp_ells_xiE} we plot the quantity $l(l+1)\zeta^E_l(r,r)/2\pi$, which is approximately the contribution to the LSS transverse velocity variance per $\log l$. We use Equation~\eqref{eq:zetaE} with the linear theory matter power spectrum computed with \textsc{CAMB}~\citep{2000ApJ...538..473L}, and focus on $r \lesssim 300 \, \mathrm{Mpc}/h$ in anticipation of the relevant distances for measuring secular parallax. Figure~\ref{fig:vperp_ells_xiE} also shows the noise power for our Gaia-like reference survey, which we introduce in Section~\ref{sec:gums}.

The linear variance of the transverse velocity field varies little over the distance range $0 \lesssim r \lesssim 300 \, \mathrm{Mpc}/h$, maintaining a roughly constant r.m.s. value of $\sigma_{v_\perp} \approx 440 \, \mathrm{km} \, \mathrm{s}^{-1}$. The primary effect of increasing the distance in Figure~\ref{fig:vperp_ells_xiE} is then to push the angular structure to smaller scales, keeping the area under each curve constant. When the radial distance is much less than the coherence scale of the velocity field, most of the power is at $l=1$, which follows from projecting a constant velocity onto the sphere (see also Section~\ref{subsec:dipoles}). As we increase the radial distance, uncorrelated patches of the velocity field come into view which project to give enhanced angular structure, and the peak of the power spectrum shifts to smaller angular scales. We also see that baryon acoustic oscillations (BAOs) in the power spectrum become visible in projection for $r \gtrsim 100 \, \mathrm{Mpc}/h$. 

\begin{figure}
  \includegraphics[width=\columnwidth]{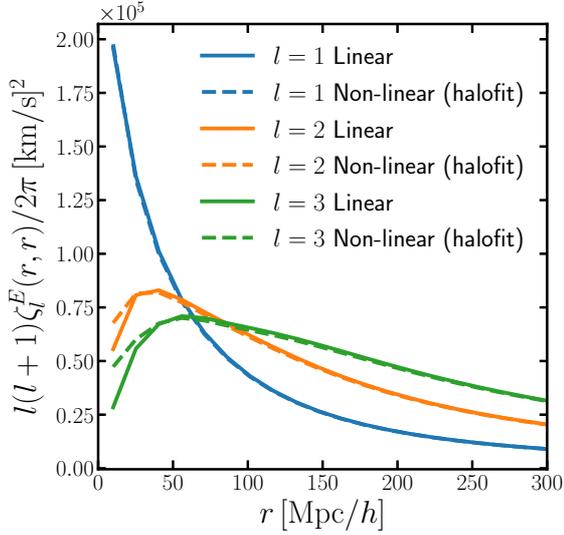}
  \caption{The first few multipoles of the LSS transverse velocity power spectrum as a function of distance. Curves ordered bottom to top at $r=300 \, \mathrm{Mpc}/h$ are $l=1$ (blue), $l=2$ (orange) and $l=3$ (green). We plot curves for both linear theory (solid) and a simplistic non-linear model which replaces the linear matter power spectrum with the \textsc{halofit} model of~\citet{2012ApJ...761..152T} (dashed).}
  \label{fig:vperp_l1l2l3_rs_xiE}
\end{figure}

To study the radial-dependence of the power spectrum more closely, in Figure~\ref{fig:vperp_l1l2l3_rs_xiE} we plot the power of the first few multipoles as a function of distance. The solid curves in this figure are the linear theory predictions, and demonstrate that the dipole power decreases strongly with distance, due to the transfer of power to smaller scales at roughly fixed total variance. Similarly, power in the higher multipoles initially increases with distance, but then falls as incoherent patches of transverse velocity become visible on the sky. Note that the proper motion due to LSS carries an additional factor of $1/r$. 

The dashed curves in Figure~\ref{fig:vperp_l1l2l3_rs_xiE} are the power spectrum computed with the non-linear correction to the matter power spectrum from \textsc{halofit}~\citep{2012ApJ...761..152T}. While this is not an accurate model for the non-linear velocity power spectrum as it does not account for non-linearity in the continuity equation, it gives us a rough idea of the sensitivity to non-linear scales. We model the non-linear unequal-time power spectrum as $P(k; r_1, r_2) = \sqrt{P(k;r_1,r_1)P(k;r_2,r_2)}$, which holds in the linear regime, and apply the \textsc{halofit} correction to the equal-time power spectra. From the figure we see that the effects of non-linear evolution are only important for nearby objects and multipoles $l \gtrsim 2$, since a small spatial scale contributes to these large angular scales only when nearby. In particular, the dipole of LSS transverse velocity is very insensitive to changing the matter power spectrum on non-linear scales\footnote{Very non-linear structures such as Virgo could still have an impact on the low-$l$ moments, since our quasi-linear approach is likely inaccurate in this case. Constrained N-body simulations may be required to accurately assess the impact of these, which we defer to a future work.}.

\begin{figure}
  \includegraphics[width=\columnwidth]{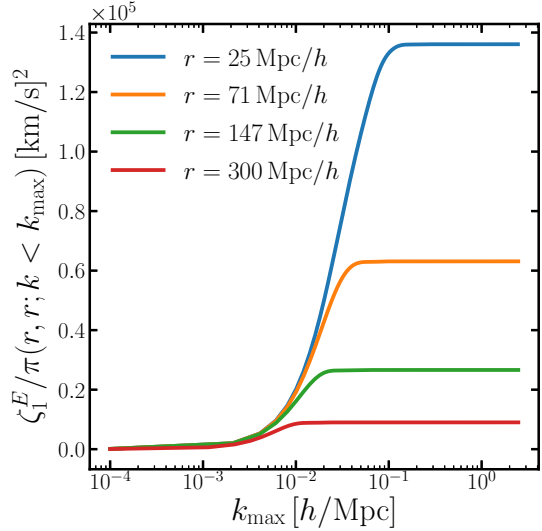}
  \caption{The cumulative contribution to the $l=1$ multipole of the LSS transverse velocity power spectrum for increasing maximum wavenumbers $k_{\mathrm{max}}$. We show curves for (top to bottom) $r = 25 \, \mathrm{Mpc}/h$ (blue), $r = 71 \, \mathrm{Mpc}/h$ (orange), $r = 147 \, \mathrm{Mpc}/h$ (green), and $r = 300 \, \mathrm{Mpc}/h$ (red). The dipole receives more contribution from small spatial scales when the LSS tracers are closer, a consequence of angular projection.}
  \label{fig:vperp_cuminteg_i1i5i10i20}
\end{figure}

This behaviour can be understood more readily in Figure~\ref{fig:vperp_cuminteg_i1i5i10i20}, where we plot the cumulative contribution to the $l=1$ power spectrum as a function of the maximum wavenumber included in the integration in Equation~\eqref{eq:zetaE}. At $z\approx0$, modifications to the matter power spectrum start to reach the percent level around $k \gtrsim 0.1 \, h\, \mathrm{Mpc}^{-1}$, but only for very nearby galaxies do these scales contribute to the dipole power, as evidenced by the convergence of the curves as $k_{\mathrm{max}}$ is increased. The forecasts in this paper include only galaxies at $r \gtrsim 20 \mathrm{Mpc}/h$, for which the contribution from non-linear scales to the dipole is negligible; the transverse velocity dipole from objects at $r \gtrsim 20 \mathrm{Mpc}/h$ is primarily sensitive to large, linear scales. This is increasingly true at greater distances due to angular projection. 

\begin{figure}
  \includegraphics[width=\columnwidth]{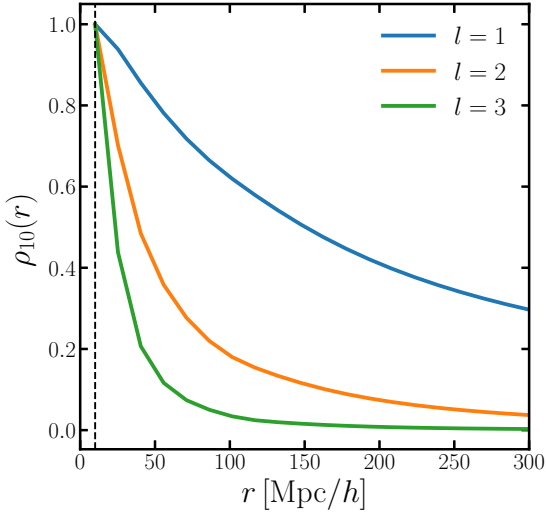}
  \caption{Dimensionless correlation coefficient of the LSS transverse velocity power spectrum between different radii, normalized at $r = 10 \, \mathrm{Mpc}/h$ (indicated by the dashed vertical line). The first few multipoles are shown as the solid curves at (top to bottom) $l=1$ (blue), $l=2$ (orange) and $l=3$ (green). Higher multipoles decorrelate with distance faster than lower multipoles due to the smaller-scale structures typically probed by higher multipoles.}
  \label{fig:vperp_l1l2l3_rs_xiE_corr}
\end{figure}

By studying the off-diagonal structure of $\zeta^E_l(r_1,r_2)$ we can study the coherence of the transverse velocity field. In Figure~\ref{fig:vperp_l1l2l3_rs_xiE_corr} we plot the dimensionless correlation coefficient for the first few multipoles normalized at $10 \, \mathrm{Mpc}/h$,  defined as $\rho_{10}(r) =  \zeta^E_l(r_{10},r)/\sqrt{\zeta^E_l(r_{10},r_{10})\zeta^E_l(r,r)}$ with $r_{10} = 10 \, \mathrm{Mpc}/h$. The radial correlation length $r_{\mathrm{co}}$ can then be defined for each $l$ as the distance where $\rho_{10}$ drops below a specified value, for example $\rho_{10}(r_{\mathrm{co}}) = 0.1$. This figure shows that the correlation length is larger on larger angular scales, which makes intuitive sense since the angular power at lower $l$ is primarily sensitive to larger spatial scales, as evidenced by the difference between the solid and dashed curves in Figure~\ref{fig:vperp_l1l2l3_rs_xiE}. Summing over all multipoles we find the coherence length of the linear transverse velocity field at $z\approx 0$ is roughly $200 \, \mathrm{Mpc}/h$, consistent with that of the three-dimensional linear velocity field~\citep{2013PhRvD..88j3510Z}.

\begin{figure}
  \includegraphics[width=\columnwidth]{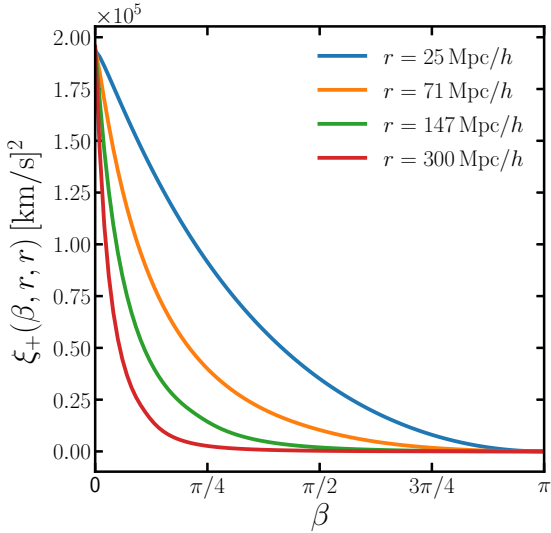}
  \caption{The $\xi_+$ correlation function of LSS transverse velocities as a function of angular separation in radians. We show curves for (top to bottom) $r = 25 \, \mathrm{Mpc}/h$ (blue), $r = 71 \, \mathrm{Mpc}/h$ (orange), $r = 147 \, \mathrm{Mpc}/h$ (green), and $r = 300 \, \mathrm{Mpc}/h$ (red). The transverse velocity field develops greater angular structure as the distance is increased, consistent with Figure~\ref{fig:vperp_ells_xiE}.}
  \label{fig:vperp_etap}
\end{figure}

\begin{figure}
  \includegraphics[width=\columnwidth]{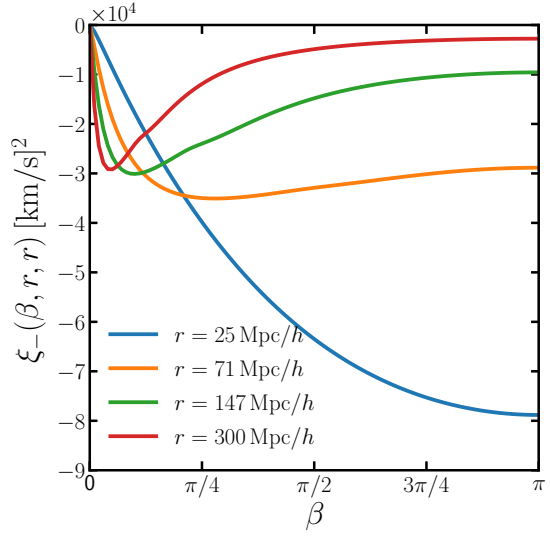}
  \caption{The $\xi_-$ correlation function of LSS transverse velocities as a function of angular separation in radians. We show curves for (bottom to top at $\beta = \pi$) $r = 25 \, \mathrm{Mpc}/h$ (blue), $r = 71 \, \mathrm{Mpc}/h$ (orange), $r = 147 \, \mathrm{Mpc}/h$ (green), and $r = 300 \, \mathrm{Mpc}/h$ (red). The transverse velocity field develops greater angular structure as the distance is increased, consistent with Figure~\ref{fig:vperp_ells_xiE}.}
  \label{fig:vperp_etam}
\end{figure}

In Figures~\ref{fig:vperp_etap} and~\ref{fig:vperp_etam} we plot the $\xi_+$ and $\xi_-$ correlation functions respectively as a function of angular separation, for different radial distances. The dependence on angular separation in $\xi_+$ reflects the $l$-dependence of the power spectrum in Figure~\ref{fig:vperp_ells_xiE}; at low radii most of the signal is dipolar, and so the transverse velocity is correlated over large angles. As we go to higher radii the power spectrum develops structure on smaller angular scales and most of the support of the correlation functions is for small $\beta$. The shape of the $\xi_-$ correlation functions can be explained by similar reasoning, albeit with the added complication that all curves must go to zero at vanishing separation due to the two components of transverse velocity being uncorrelated at a point.
%
%

Similarly to the radial coherence distance, we can define a coherence angle as the angular separation at which the correlation function $\xi_+$ falls below 10\% of its zero-lag value. We find that even at the largest distance we include in our forecasts ($r \sim 300 \, \mathrm{Mpc}/h$) the angular coherence scale is large at roughly $20^\circ$.

%
We close this section by noting that the LSS correlation functions may be computed more directly by using the explicit forms of the Wigner $d$ functions and the identity between spherical Bessel functions given in Equation~\eqref{appeq:greateq} to perform the sum over multipoles. We find that
\begin{align}
  \xi_\pm(\beta_{12},r_1,r_2) &= \int \frac{k^2\mathrm{d}k}{2\pi^2} \, P_v(k;r_1,r_2) \nonumber \\
  & \, \times \left[ (\pm1 + \cos \beta_{12})\frac{j_1(kx)}{kx} - \sin^2 \beta_{12} \frac{(kr_1)(kr_2)}{(kx)^2}j_2(kx) \right],
\end{align}
where $P_v(k;r_1,r_2)$ is the velocity potential power spectrum and $x = (r_1^2 + r_2^2 - 2r_1r_2\cos\beta_{12})^{1/2}$ is the three-dimensional distance between the two points. Taking the limit $\beta_{12} \rightarrow 0$ with $r_1 = r_2$ gives $\xi_+ \rightarrow 2\sigma_v^2/3$, i.e. the transverse velocity carries two thirds of the total velocity variance at a point, with the radial velocity carrying the remaining third, as required by isotropy.

\section{Likelihood formalism}
\label{sec:like}

In this section we develop the likelihood formalism which will allow us to derive optimal estimators for various sources of extragalactic proper motion. We focus on proper motions from the secular parallax due to the relative motion of the Solar System barycentre with respect to galaxies and quasars, secular aberration drift due to the time-varying aberration caused by the acceleration of the Solar System barycentre towards the galactic centre, and the intrinsic peculiar motion of galaxies and quasars due to large-scale structure.

\subsection{Uncertainty from noise and peculiar velocities}
\label{subsec:ncov}

Suppose we observe a set of $N$ objects and measure the complex proper motion ${}_+d$ and ${}_-d$ (c.f. Equation~\eqref{eq:vperp_rotate}) for each object. Define the data vector $\mathbf{d} = ({}_+\mathbf{d}^\intercal, {}_-\mathbf{d}^\intercal)^\intercal$, where ${}_\pm\mathbf{d}$ are vectors containing the measurements from each object. Assuming the measurement noise is uncorrelated between objects, the noise covariance matrix is given by
\begin{align}
  \mathbfss{N} &= \begin{pmatrix}
    \mathbfss{N}_+ & \mathbfss{N}_- \\
    \mathbfss{N}_-^\dagger & \mathbfss{N}_+^* \end{pmatrix}, \nonumber \\
  &= \begin{pmatrix}
    \mathrm{diag}[\sigma_{\mu \delta}^{(i)2} + \sigma_{\mu \alpha^*}^{(i)2}] & \mathrm{diag}[\sigma_{\mu \delta}^{(i)2} - \sigma_{\mu \alpha^*}^{(i)2}] \\
    \mathrm{diag}[\sigma_{\mu \delta}^{(i)2} - \sigma_{\mu \alpha^*}^{(i)2}] & \mathrm{diag}[\sigma_{\mu \delta}^{(i)2} + \sigma_{\mu \alpha^*}^{(i)2}] \end{pmatrix},
\end{align}
where $\sigma_{\mu \delta}^{(i)2}$ and $\sigma_{\mu \alpha^*}^{(i)2}$ are the proper motion noise variances in the declination and corrected-right-ascension (i.e. true arc) directions respectively for the $i^{\mathrm{th}}$ object (throughout this section, $\mu$ labels proper motions). The diagonal matrices in each block are $N \times N$, with the full noise covariance matrix a $2N \times 2N$ dimensional matrix. For isotropic noise we would have $\sigma_{\mu \delta}^{(i)2} = \sigma_{\mu \alpha^*}^{(i)2} \equiv \sigma_\mu^{(i)2}$, meaning the covariance matrix is diagonal with $\mathbfss{N} = 2(\sigma_\mu^{2(i)} \mathbfss{I}_N) \oplus (\sigma_\mu^{2(i)} \mathbfss{I}_N)$, but in general these noise variances will be different due to the scanning strategy. The noise variance will also differ for each object due to differences in magnitude, colour, and other intrinsic galaxy or quasar properties.

The mean of the data vector over the noise has contributions from the secular parallax velocity, the secular aberration drift proper motion, and the peculiar velocity due to LSS. We further split the LSS peculiar velocity into a part correlated with the Solar System's motion relative to the CMB (i.e. the SP velocity), and an uncorrelated part; a correlated part is expected since part of the SP velocity is due to the Milky Way's velocity relative to the CMB, $\mathbf{V}_g$, which is correlated with that of nearby galaxies due to the gravitational effects of large-scale structure. We write $\mathbf{v}^{\mathrm{LSS}} = \mathbf{v}_c^{\mathrm{LSS}} + \mathbf{v}_u^{\mathrm{LSS}}$, where the correlated part is given by
\begin{equation}
  \mathbf{v}_c^{\mathrm{LSS}} = \langle \mathbf{v}_c^{\mathrm{LSS}} \mathbf{v}_c^{\intercal \mathrm{SP}} \rangle \langle  \mathbf{v}_c^{\mathrm{SP}}  \mathbf{v}_c^{\intercal \mathrm{SP}} \rangle^{-1}  \mathbf{v}_c^{\mathrm{SP}},
\end{equation}
where $\mathbf{v}_c^{\mathrm{SP}}$ is the part of the SP velocity correlated with large-scale structure. We take this to be (minus) the Milky Way-CMB relative velocity vector, whose magnitude is roughly $550 \, \mathrm{km/s}$ towards galactic coordinates $(l,b) = (266.5^\circ, 29.1^\circ)$~\citep{1993ApJ...419....1K}. By isotropy the quantity $\langle  \mathbf{v}_c^{\mathrm{SP}}  \mathbf{v}_c^{\intercal \mathrm{SP}} \rangle$ must be proportional to $\delta_{ab}$ which implies that $\langle  \mathbf{v}_c^{\mathrm{SP}}  \mathbf{v}_c^{\intercal \mathrm{SP}} \rangle = (\sigma_{\mathrm{SP},c}^2/3) \delta_{ab}$ where $\sigma_{\mathrm{SP},c}^2$ is the variance of the correlated part of the Solar System's velocity, equal to the variance of the large-scale structure velocity field $\sigma_v^2$ at our location (and by statistical homogeneity, any location on the $z=0$ time slice). The covariance of the correlated LSS velocity at spatial location $(\hat{\mathbf{n}}, r)$  with the correlated SP velocity is given by $\langle \mathbf{v}_c^{\mathrm{LSS}} \mathbf{v}_c^{\intercal \mathrm{SP}} \rangle  = \langle \mathbf{v}(\hat{\mathbf{n}}, r) \mathbf{v}^\intercal(\mathbf{r} = 0) \rangle $. As shown in~\citet{1988ApJ...332L...7G}, this may be written as $ \langle v_a(\hat{\mathbf{n}}, r) v_b(\mathbf{r} = 0) \rangle = \Psi_\perp(r) \delta_{ab} + \left[\Psi_\parallel(r) - \Psi_\perp(r) \right]\hat{n}_a\hat{n}_b$. After projecting transverse to the line-of-sight, we find the part of the complex proper motion due to LSS correlated with the local SP velocity is
\begin{equation}
  {}_\pm \boldsymbol{\mu}_c^{\mathrm{LSS}} = -\frac{3}{\sigma_v^2}\Psi_\perp(r){}_\pm \boldsymbol{\mu}_c^{\mathrm{SP}},
  \label{eq:muclss}
\end{equation}
where~\citep{1988ApJ...332L...7G}
\begin{equation}
  \Psi_\perp(r) = \int \frac{k^2 \mathrm{d}k}{2\pi^2} P_v(k;r) \frac{j_1(kr)}{kr},
\end{equation}
with $P_v(k;r)$ the velocity power spectrum. Note that as $r \rightarrow 0$ we have $\Psi_\perp(r) \rightarrow \sigma^2_v/3$, which implies that $ {}_\pm \boldsymbol{\mu}_c^{\mathrm{LSS}} \rightarrow -{}_\pm \boldsymbol{\mu}_c^{\mathrm{SP}}$, i.e. the total observed proper motion averaged over galaxies tends to the part of the SP velocity uncorrelated with LSS, which we take to be the velocity of the Solar System relative to that of the Milky Way.

\begin{figure}
  \includegraphics[width=\columnwidth]{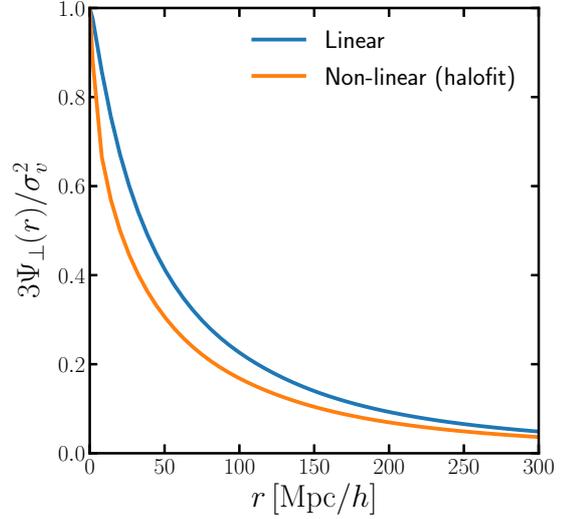}
  \caption{The perpendicular velocity correlation function $\Psi_\perp(r)$ of~\citet{1988ApJ...332L...7G} normalized by the one-dimensional variance $\sigma_v^2/3$, in linear theory (blue, upper curve), and with the \textsc{halofit} correction to the matter power spectrum (orange, lower curve).}
  \label{fig:gorski_perp_normed}
\end{figure}

In Figure~\ref{fig:gorski_perp_normed} we plot the quantity $3\Psi_\perp(r)/\sigma^2_v$ with both the linear velocity power spectrum (upper curve) and the velocity power spectrum using the \textsc{halofit} correction to the matter power spectrum but assuming the linear continuity equation (lower curve) to give a rough idea of the contribution from non-linear scales. From this figure we see that the perpendicular correlation function falls to 10\% of its zero-lag value by roughly $200 \, \mathrm{Mpc}/h$, and that the contribution from non-linear scales can be quite large for distances less than this. This quantity is also sensitive to the chosen cosmological parameters, which impact the velocity power spectrum through the growth factor, the Hubble parameter, and the shape of the matter spectrum.

The remaining part of the peculiar velocity due to LSS is uncorrelated with our local velocity. This part is zero on average, with a covariance matrix $\mathbfss{C}_{\pm}^u \equiv \langle {}_+\boldsymbol{\mu}^{\mathrm{LSS}}_u  {}_\mp\boldsymbol{\mu}^{\intercal \mathrm{LSS}}_u \rangle$ given by
\begin{equation}
  C_{\pm,ij}^u = C_{\pm,ij}^{\mathrm{LSS}} - \frac{3\Psi_\perp(r_i)}{\sigma_v^2}\frac{3\Psi_\perp(r_j)}{\sigma_v^2} \langle {}_+\mu^{\mathrm{SP}}_{c,i}  {}_\mp\mu^{\mathrm{SP}}_{c,j} \rangle,
\end{equation}
where $C_{\pm,ij}^{\mathrm{LSS}}$ are elements of the full LSS complex proper motion covariance matrix. Since the local SP velocity is purely dipolar, the E-mode power spectrum of the uncorrelated LSS transverse velocity is given by
\begin{equation}
  \zeta_l^{E,u}(r_1, r_2) = \zeta_l^E(r_1, r_2) - \frac{8\pi}{3}\frac{3\Psi_\perp(r_i)}{\sigma_v^2}\frac{3\Psi_\perp(r_j)}{\sigma_v^2} \frac{\sigma_v^2}{3}\delta_{l1}.
\end{equation}
In the limit that $r_1$ and $r_2$ both tend to zero we have $\zeta_l^{E,u} \rightarrow 0$, since in this limit the proper motion due to LSS is completely correlated with the velocity of the Milky Way.

We now marginalize over the uncorrelated proper motion due to LSS, assuming it obeys Gaussian statistics\footnote{Non-Gaussianity could be incorporated with constrained realizations of a Milky Way-like local environment.} with covariance matrix $\mathbfss{C}_{\pm}^u$\footnote{Any quasar intrinsic motion due to radio jet variability, assuming it obeys Gaussian statistics, could be marginalized over and included in the noise variance since it is uncorrelated between different objects. We focus on galaxies in this work, and do not include this additional noise in the small number of quasars included in our forecasts.}. Assuming Gaussian noise, the likelihood of the data vector becomes, up to an arbitrary constant,
\begin{align}
  &-2\ln p(\mathbf{d} |  {}_\pm \boldsymbol{\mu}^{\mathrm{SP}} , {}_\pm \boldsymbol{\mu}^{\mathrm{SAD}}) =  \nonumber\\
  & \, \begin{pmatrix}
    {}_+\mathbf{d} - {}_{+}\boldsymbol{\mu}\\
    {}_-\mathbf{d} - {}_{-}\boldsymbol{\mu} \end{pmatrix}^\dagger
  \begin{pmatrix}
    \mathbfss{N}_+ + \mathbfss{C}_+^u& \mathbfss{N}_- + \mathbfss{C}_-^u\\
    (\mathbfss{N}_- + \mathbfss{C}_-^u)^\dagger & (\mathbfss{N}_+ + \mathbfss{C}_+^u)^* \end{pmatrix}^{-1}
  \begin{pmatrix}
    {}_+\mathbf{d} - {}_{+}\boldsymbol{\mu}\\\
    {}_-\mathbf{d} - {}_{-}\boldsymbol{\mu}  \end{pmatrix},
  \label{eq:noiselike}
\end{align}
where
\begin{equation}
   {}_{\pm}\boldsymbol{\mu} \equiv {}_\pm\boldsymbol{\mu}^{\mathrm{SP}} + {}_\pm\boldsymbol{\mu}^{\mathrm{SAD}} + {}_\pm\boldsymbol{\mu}^{\mathrm{LSS}}_c,
\end{equation}
with ${}_\pm\boldsymbol{\mu}^{\mathrm{LSS}}_c$ given by Equation~\eqref{eq:muclss}. Note that $(\mathbfss{N}_+ + \mathbfss{C}_+^u) = (\mathbfss{N}_+ + \mathbfss{C}_+^u)^\dagger$ and $(\mathbfss{N}_- + \mathbfss{C}_-^u) = (\mathbfss{N}_- + \mathbfss{C}_-^u)^\intercal$. Equation~\eqref{eq:noiselike} is a Gaussian likelihood function for the observed proper motion, conditioned on the Solar System's velocity relative to the CMB, its acceleration towards to the galactic centre, and the Milky Way's velocity relative to the CMB. This function contains all the information available under the assumption that both sources of noise (instrumental and cosmic variance) obey Gaussian statistics.

\begin{figure*}
  \includegraphics[width=1.5\columnwidth]{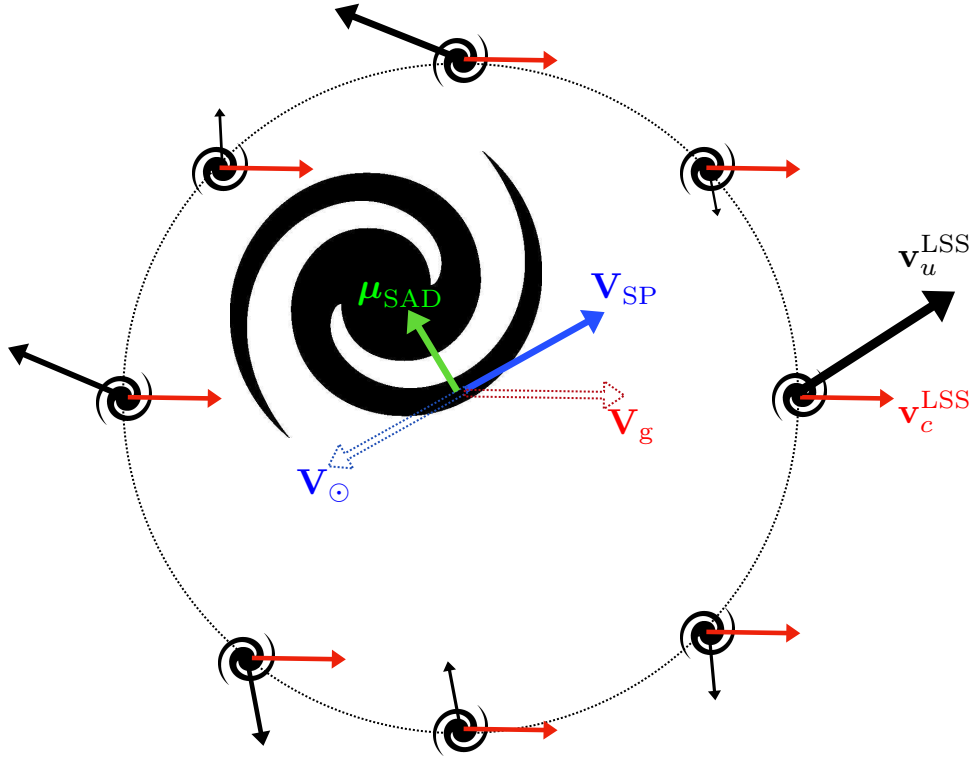}
  \caption{Schematic diagram of the various sources of proper motion considered in this work, in the CMB rest frame. The Sun is at the centre, and has a velocity $\mathbf{V}_{\sun}$ indicated by the dashed blue arrow. This induces the SP velocity vector $\mathbf{V}_{\mathrm{SP}}$, indicated by the solid blue arrow, which gives rise to a dipolar proper motion in each surrounding galaxy; the solid blue vector should be added to each galaxy and projected perpendicular to $\hat{\mathbf{n}}$ to find the SP proper motion. The SAD proper motion vector is indicated by the solid green arrow, which is directed towards the Milky Way centre (the large central galaxy); again, this solid vector should be added to each surrounding galaxy to find its SAD proper motion. In addition, each galaxy has a peculiar velocity, which consists of a random component $\mathbf{v}^{\mathrm{LSS}}_u$ (black solid arrows) and a part $\mathbf{v}^{\mathrm{LSS}}_c$ correlated with the motion of the Milky Way (red solid arrows). The Milky Way velocity $\mathbf{V}_g$ is indicated by the red dashed arrow. All galaxies are here assumed to be at the same radial distance, indicated by the dashed circle, so the lengths of the red solid vectors are equal; in general they are given by Equation~\eqref{eq:muclss}. The total proper motion of each object is found by summing up the solid arrows centred on that object, as well as the two solid arrows centred on the observer. The average of these proper motions has a contribution proportional to $\mathbf{V}_g$.}
  \label{fig:vel_diagram}
\end{figure*}

In Figure~\ref{fig:vel_diagram} we give a schematic diagram which illustrates the various contributions to the proper motion of each galaxy.

\subsection{Optimal estimators for large-scale proper motion}
\label{subsec:estimators}
  
The SP and SAD complex proper motions of object $i$ at radial distance $r_i$ are given by
\begin{align}
  {}_\pm\mu^{\mathrm{SP}}_i &= \mathbf{V}^{\mathrm{SP}} \cdot [\hat{\boldsymbol{\theta}}^{(i)} \pm i \hat{\boldsymbol{\phi}}^{(i)}]/r_i \equiv \left[\mathbfss{S}^{(1)}_\pm \mathbf{V}^{\mathrm{SP}} \right]_i\\
  {}_\pm\mu^{\mathrm{SAD}}_i &= \boldsymbol{\mu}^{\mathrm{SAD}} \cdot [\hat{\boldsymbol{\theta}}^{(i)} \pm i \hat{\boldsymbol{\phi}}^{(i)}] \equiv \left[\mathbfss{S}^{(0)}_\pm \boldsymbol{\mu}^{\mathrm{SAD}} \right]_i,
\end{align}
where $\mathbf{V}^{\mathrm{SP}} $ is the fixed three-dimensional SP velocity vector, i.e. minus the Solar System barycentre's velocity with respect to the CMB rest frame, $\boldsymbol{\mu}^{\mathrm{SAD}}$ is a three-dimensional `acceleration' vector having units of proper motion and parallel to the vector connecting the observer to the galactic centre, and we have introduced the $N \times 3$ matrix $\mathbfss{S}^{(n)}_\pm$ whose elements are given by $S^{(n)}_{\pm,ia} = (\hat{\theta}^{(i)}_a \pm i\hat{\phi}^{(i)}_a)/r_i^n$. Note that the SAD proper motion is independent of distance. Inserting these expressions into Equation~\eqref{eq:noiselike} we find, after some manipulation
\begin{equation}
  -2\ln p(\mathbf{d} |\boldsymbol{\mu}^{\mathrm{SAD}}, \mathbf{V}^{\mathrm{SP}}) = \begin{pmatrix}
    \Delta \boldsymbol{\mu}^{\mathrm{SAD}} \\
    \Delta \mathbf{V}^{\mathrm{SP}} \end{pmatrix}^\intercal \begin{pmatrix}
    \mathbfss{M}^{(0)} & \mathbfss{M}^{(1)} \\
    \mathbfss{M}^{(1)\intercal} & \mathbfss{M}^{(2)} \end{pmatrix}
  \begin{pmatrix}
    \Delta \boldsymbol{\mu}^{\mathrm{SAD}} \\
    \Delta \mathbf{V}^{\mathrm{SP}} \end{pmatrix},
  \label{eq:vmulike}
\end{equation}
up to an arbitrary constant. Denoting the total covariance matrix $\mathbfss{C}_T \equiv \mathbfss{N} + \mathbfss{C}^u$, the $3\times 3$ matrices $\mathbfss{M}^{(n)}$ are given by
\begin{align}
  \mathbfss{M}^{(0)} &\equiv 2\mathbfss{S}_+^{(0)\dagger}(\mathbfss{C}_T^{-1})_{(1,1)} \mathbfss{S}_+^{(0)} + 2\mathrm{Re}[ \mathbfss{S}_+^{(0)\dagger}(\mathbfss{C}_T^{-1})_{(1,2)} \mathbfss{S}_-^{(0)}]\nonumber ,\\
  \mathbfss{M}^{(1)} &\equiv 2\mathbfss{S}_+^{(0)\dagger}(\mathbfss{C}_T^{-1})_{(1,1)} \mathbfss{S}_+^{(1)} + 2\mathrm{Re}[ \mathbfss{S}_+^{(0)\dagger}(\mathbfss{C}_T^{-1})_{(1,2)} \mathbfss{S}_-^{(1)}]\nonumber ,\\
    \mathbfss{M}^{(2)} &\equiv 2\mathbfss{S}_+^{(1)\dagger}(\mathbfss{C}_T^{-1})_{(1,1)} \mathbfss{S}_+^{(1)} + 2\mathrm{Re}[ \mathbfss{S}_+^{(1)\dagger}(\mathbfss{C}_T^{-1})_{(1,2)} \mathbfss{S}_-^{(1)}],
  \label{eq:Mmatrices}
\end{align}
where $(\mathbfss{C}_T^{-1})_{(1,1)}$ denotes the $(+,+)$ block of the inverse covariance matrix, and similarly for the other blocks. In the limit of no cosmic variance from LSS, these matrices reduce to
\begin{equation}
  \mathsf{M}^{(n)}_{ab}\bigr\rvert_{\mathrm{\mathrm{LSS}} = 0} = \sum_{i=1}^N \left( \frac{\hat{\theta}^{(i)}_a\hat{\theta}^{(i)}_b}{\sigma_{\mu \delta}^{(i)2}} +  \frac{\hat{\phi}^{(i)}_a\hat{\phi}^{(i)}_b}{\sigma_{\mu \alpha^*}^{(i)2}} \right)r_i^{-n}.
  \label{eq:Mnoise}
\end{equation}
The means are given by $\Delta \boldsymbol{\mu}^{\mathrm{SAD}} = \boldsymbol{\mu}^{\mathrm{SAD}} - \hat{\boldsymbol{\mu}}^{\mathrm{SAD}}$ and $\Delta \mathbf{V}^{\mathrm{SP}} = \mathbf{V}^{\mathrm{SP}} - \hat{\mathbf{V}}^{\mathrm{SP}}$\footnote{Note that hats on quantities now refer to estimators rather than directions, as should be clear from the context.} with
\begin{align}
  \hat{\mathbf{V}}^{\mathrm{SP}} &= \left[\mathbfss{M}^{(2)} - \mathbfss{M}_S^{(1)}\mathbfss{M}^{(0)-1}\mathbfss{M}^{(1)}_S\right]^{-1}\left[\mathbf{L}^{(1)} - \mathbfss{M}_S^{(1)}\mathbfss{M}^{(0)-1}\mathbf{L}^{(0)}\right], \label{eq:vsphat} \\
  \hat{\boldsymbol{\mu}}^{\mathrm{SAD}} &= \left[\mathbfss{M}^{(0)} - \mathbfss{M}_S^{(1)}\mathbfss{M}^{(2)-1}\mathbfss{M}_S^{(1)}\right]^{-1}\left[\mathbf{L}^{(0)} - \mathbfss{M}_S^{(1)}\mathbfss{M}^{(2)-1}\mathbf{L}^{(1)}\right], \label{eq:musadhat}
\end{align}
where $\mathbfss{M}^{(1)}_S$ is the symmetric part of $\mathbfss{M}^{(1)}$ and the three-vectors $\mathbf{L}^{(n)}$ are given by
\begin{equation}
  \mathbf{L}^{(n)} = 2\mathrm{Re}\left\{ \left[ \mathbfss{S}_+^{(n)\dagger}(\mathbfss{C}_T^{-1})_{(1,1)} + \mathbfss{S}_-^{(n)\dagger}(\mathbfss{C}_T^{-1})^*_{(1,2)} \right] \left ({}_+\mathbf{d} - {}_+\boldsymbol{\mu}^{\mathrm{LSS}}_c\right)\right\}.
\end{equation}
In the limit of no cosmic variance, these vectors reduce to
\begin{equation}
  \mathsf{L}^{(n)}_a\bigr\rvert_{\mathrm{\mathrm{LSS}} = 0} = \sum_{i=1}^N \left( \frac{d_{\hat{\theta}}^{(i)}\hat{\theta}^{(i)}_a}{\sigma_{\mu \delta}^{(i)2}} + \frac{d_{\hat{\phi}}^{(i)}\hat{\phi}^{(i)}_a}{\sigma_{\mu \alpha^*}^{(i)2}} \right)r_i^{-n},
  \label{eq:Lnoise}
\end{equation}
where $d_{\hat{\theta}}$ and $d_{\hat{\phi}}$ are the declination and right-ascension components of the measured proper motion respectively, with the correlated LSS part subtracted off.

\subsection{Bias and variance of the proper motion estimators}
\label{subsec:bias}

The estimators for the three-vectors given in Equations~\eqref{eq:vsphat} and~\eqref{eq:musadhat} are optimal estimators for SP velocity and SAD acceleration vectors. If the correlated part of the LSS velocity were known perfectly, these estimators would be unbiased, with $\langle \hat{\mathbf{V}}^{\mathrm{SP}} \rangle = \mathbf{V}^{\mathrm{SP}}$ and similarly for $\boldsymbol{\mu}^{\mathrm{SAD}}$. After subtraction of the correlated LSS term, the vectors $\mathbf{L}^{(n)}$ have the form of correlations of the inverse-variance weighted data vector with the spherical polar basis vectors, weighted with the appropriate factor of $r$, then summed over the $N$ objects in the proper motion catalogue.

In general, the correlated term $\boldsymbol{\mu}^{\mathrm{LSS}}_c$ will not be known perfectly, due to the uncertain form of $3\Psi_\perp(r)/\sigma_v^2$ from non-linearity (see Figure~\ref{fig:gorski_perp_normed}) and uncertainty in the cosmological parameters. An alternative way of treating this term is to ignore it in the estimator and study the size of the resulting \emph{bias} in the mean of $ \hat{\mathbf{V}}^{\mathrm{SP}}$ and $\hat{\boldsymbol{\mu}}^{\mathrm{SAD}}$. In a similar vein, we could consider neglecting the cosmic variance due to uncorrelated LSS proper motion altogether, and just adopt the instrument-noise-only expressions of Equation~\eqref{eq:Mnoise} and Equation~\eqref{eq:Lnoise}. This is necessary when the number of objects in the proper motion catalogue is large enough that the matrix inversion of $\mathbfss{C}_T$ (a non-diagonal matrix) is slow, but small enough that the large-$N$ approximations made below are inaccurate. In this case, the uncorrelated LSS proper motion simply adds noise to the estimator.

The covariance matrix between the SAP and SP estimators can be read off from Equation~\eqref{eq:vmulike}, since both estimators are Gaussian distributed. Note that in the limit that $\mathbf{V}^{\mathrm{SP}}$ is perfectly known, the optimal estimator and variance of $\boldsymbol{\mu}^{\mathrm{SAD}}$ may be found by taking $\mathbfss{M}^{(2)} \rightarrow \infty$ at fixed $\mathbfss{M}^{(0)}$, and conversely if $\boldsymbol{\mu}^{\mathrm{SAD}}$ is perfectly known we can take $\mathbfss{M}^{(0)} \rightarrow \infty$ at fixed $\mathbfss{M}^{(2)}$.

Finally, it is straightforward to show that the covariance matrix of the $\mathbf{L}$ vectors is given by
\begin{equation}
  \langle \Delta \mathbf{L}^{(m)} \Delta \mathbf{L}^{(n)\intercal} \rangle = \mathbfss{M}^{(m+n)}.
  \label{eq:Lcov}
\end{equation}
In the scenario where we treat cosmic variance as `extra noise' and use the instrument-noise-only estimators, Equation~\eqref{eq:Lcov} receives additional contributions from uncorrelated LSS which are straightforward to compute in our formalism.

The $\mathbfss{M}$ matrices and $\mathbf{L}$ vectors are the fundamental building blocks which we must compute before we form the optimal estimators for the SP and SAD three-vectors in Equations~\eqref{eq:vsphat} and~\eqref{eq:musadhat}. 

\subsection{Amplitude estimators and the Hubble constant}
\label{subsec:asp}

The SP velocity is known a priori from the CMB dipole, having a magnitude of $(369 \pm 0.9) \, \mathrm{km/s}$ towards galactic coordinates $(l,b) = (263.99^\circ \pm 0.14^\circ, 48.26^\circ \pm 0.03^\circ)$~\citep{2014A&A...571A..27P}. Likewise, the magnitude of the SAD proper motion is roughly $4.3 \, \mu\mathrm{as} \, \mathrm{yr}^{-1}$ towards the galactic centre~\citep{2006AJ....131.1471K, 2011A&A...529A..91T}.

Fixing the SP velocity vector to that implied by the CMB dipole\footnote{The small uncertainty in the amplitude and direction of the SP velocity vector could be incorporated with a prior added to the log-likelihood, but the errors are small enough that fixing it is sufficiently accurate for our purposes.}, we can attempt to constrain only the amplitude of the SP by parametrizing $\boldsymbol{\mu}^{\mathrm{SP}} = -A_{\mathrm{SP}}\mathbf{V_0}/r$, where $\mathbf{V_0}$ is the Solar System's motion relative to the CMB and $A_{\mathrm{SP}}$ has a fiducial value of unity. Similarly, we can attempt to constrain only the amplitude of the SAD proper motion by writing $\boldsymbol{\mu}^{\mathrm{SAD}} = A_{\mathrm{SAD}}\boldsymbol{\mu_0}$, where $A_{\mathrm{SAD}}$ has unit fiducial value. Alternatively, since the observable quantity is actually $\boldsymbol{\mu}^{\mathrm{SP}} + \boldsymbol{\mu}_c^{\mathrm{LSS}}$ we could attempt to project against the direction of this quantity; we do not take this approach since this would require knowledge of $3\Psi_\perp/\sigma_v^2$, which is contaminated by non-linearities and imperfect knowledge of the cosmological parameters. Instead we treat this extra term as a bias in $A_{\mathrm{SP}}$.

Spectroscopic redshift information provides the quantity $H_0r$\footnote{Peculiar radial velocities add noise to the distance estimate causing fluctuations in the inferred transverse velocity of an object of $\sim \frac{15\%}{(r/20 \, h^{-1}\mathrm{Mpc})}$. This noise is subdominant to other sources of variance at low distances, and is further reduced by averaging over uncorrelated patches of the peculiar radial velocity. There will also be some bias from the part of the radial velocity correlated with our local motion, but this is no more than 15\% for the objects we consider and subdominant to the bias in the transverse velocity itself. Note that averaging over both parts of the LSS transverse velocity gives no additional bias, since radial and transverse velocities are uncorrelated at a point~\citep{2012ApJ...755...58N}. Similarly, sensitivity of the FRW distance-redshift relation to the cosmological parameters is negligible at the low redshifts we consider, given the expected sensitivity of our measurements.}, so the amplitude of the SP velocity is completely degenerate with the Hubble constant $H_0$. Thus, forecast constraints on the amplitude $A_{\mathrm{SP}}$ may be interpreted as fractional constraints on the Hubble constant (c.f.~\citealt{2016A&A...589A..71B}). Alternatively we may consider the Hubble constant as fixed, in which case the amplitude of the SP effect can be constrained directly.

 Optimal estimators for $A_{\mathrm{SP}}$ and $A_{\mathrm{SAD}}$ may be straightforwardly derived from Equation~\eqref{eq:vmulike}. For example, fixing the SAD proper motion, the optimal estimator for $A_{\mathrm{SP}}$ is
 \begin{equation}
   \hat{A}_{\mathrm{SP}} = \left(\mathbf{V}_0^\intercal \mathbfss{M}^{(2)}\mathbf{V}_0\right)^{-1} \mathbf{V}_0^\intercal \mathbf{L}^{(1)},
 \end{equation}
 which has variance $\sigma_{A_{\mathrm{SP}}}^2 = \left(\mathbf{V}_0^\intercal \mathbfss{M}^{(2)}\mathbf{V}_0\right)^{-1}$.

\subsection{Large $N$, isotropic noise limit}
\label{subsec:LNIN}

In the limit where the cosmic variance from uncorrelated LSS proper motion is non-negligible, we must invert the $N\times N$ matrices~$(\mathbfss{N}_\pm + \mathbfss{C}_\pm^u)$ to form the optimal estimators in Equation~\eqref{eq:vsphat} and~\eqref{eq:musadhat}. This is clearly impractical if $N$ is large, as it will be for current astrometric surveys. Fortunately, an accurate approximation to this term exists when the instrumental noise is isotropic (which is approximately true for current astrometric surveys) and $N$ is larger than a few hundred.

Consider first the cosmic-variance-free form of the $\mathbfss{M}$ matrices given in Equation~\eqref{eq:Mnoise}. Setting $\sigma_{\mu \delta}^{(i)2} = \sigma_{\mu \alpha^*}^{(i)2} \equiv \sigma_\mu^{(i)2}$ and approximating the sum over objects with an integral over angular and radial position, we find that
\begin{equation}
  \mathsf{M}^{(n)}_{ab}\bigr\rvert_{\mathrm{\mathrm{LSS}} = 0} \approx \frac{2N}{3} \left \langle \frac{1}{\sigma_\mu^2 r^n} \right\rangle \delta_{ab},
  \label{eq:LNINMnocv}
\end{equation}
where the angle brackets denote an average over the catalogue, accounting for the potential correlations between noise variance and distance that one might expect in a flux-limited survey. Note that we have implicitly neglected any correlations between the distance and the angular positions, i.e. we assume a uniform depth across the survey\footnote{In reality there will be preferential selection of nearby objects in areas of sky where the noise variance is higher, inducing correlations between sky position and distance and mainly affecting the faintest sources just above the detection threshold. We note that in this situation the LNIN approximations made in this section are likely to be poor, but the full likelihood formalism of the previous section can still be used.}. The Kronecker delta is clearly demanded by isotropy in this scenario.

In this large $N$, isotropic noise (LNIN) limit, the variance of one component of the SP velocity after marginalizing over the SAD proper motion is, in the absence of cosmic variance,
\begin{equation}
  \sigma_{v_x}^2\bigr\rvert_{\mathrm{\mathrm{LSS}} = 0}  = \frac{3}{2N} \left(\left \langle \sigma_\mu^{-2} r^{-2} \right\rangle - \frac{\left \langle  \sigma_\mu^{-2} r^{-1} \right \rangle^2}{\left \langle  \sigma_\mu^{-2} \right \rangle}\right)^{-1}.
  \label{eq:sigvxnonoiseLNIN}
\end{equation}
Equation~\eqref{eq:sigvxnonoiseLNIN} clearly demonstrates the importance of having radial information to disentangle the effects of SP and SAD proper motion. These have the same the angular structure (E-mode dipoles), and differ only in their radial dependence. In the scenario where all the objects in the catalogue are at the same radial distance, the variance on each component of the SP velocity tends to infinity. Similarly, the variance of the SP amplitude $A_{\mathrm{SP}}$ in this scenario is simply given by $\sigma_{A_{\mathrm{SP}}}^2 = \sigma_{v_x}^2/\lvert \mathbf{V}_0 \rvert^2$.

We can use the LNIN approximation to compute the inverse of the matrices $(\mathbfss{N}_\pm + \mathbfss{C}_\pm^u)$. Firstly, we will bin all the galaxies into $N_r$ distance bins, and assume that the noise variance is constant within each bin. We then use Equation~\eqref{eq:xipmlms} to write (taking care to correct for the phase factors)
\begin{equation}
  (\mathbfss{N}_\pm + \mathbfss{C}_\pm^u)_{ij} = \sum_{lm} \left[\boldsymbol{\Omega}_l^{(\pm)}\right]_{r_i,r_j} {}_1Y_{lm}(\hat{\mathbf{n}}_i){}_{\pm 1}Y_{lm}^*(\hat{\mathbf{n}}_j),
  \label{eq:cplusn}
\end{equation}
where we have defined the quantities
\begin{align}
  \left[\boldsymbol{\Omega}_l^{(+)}\right]_{r_i, r_j} &= N(r_i)\delta_{r_i,r_j} + \zeta^{E,u}_l(r_i, r_j), \nonumber \\
  \left[\boldsymbol{\Omega}_l^{(-)}\right]_{r_i, r_j} &= -\zeta^{E,u}_l(r_i, r_j),
\end{align}
where $N(r_i)$ is the total noise power spectrum in the bin at distance $r_i$ (containing $N_{r_i}$ objects) , given by $N(r_i) = 2\sigma_\mu^2(r_i)\times 4\pi/N_{r_i}$. Note that the noise power consists of E and B modes in equal ratio. The minimum bin width should be the typical distance uncertainty caused by radial peculiar velocities (for spectroscopic redshifts), which is roughly $\Delta r \approx 3 \, \mathrm{Mpc}/h$ at $z=0$. We use a maximum of $N_r = 20$ redshift bins, with $\Delta r \approx 15 \, \mathrm{Mpc}/h$.

Having separated the radial and angular dependence of the quantities in Equation~\eqref{eq:cplusn}, we can use the completeness and orthogonality relations of the spin-weighted spherical harmonics (Equations~\eqref{appeq:orthog} and~\eqref{appeq:completeness}) to write
\begin{equation}
  (\mathbfss{N}_\pm + \mathbfss{C}_\pm^u)^{-1}_{ij} \approx \frac{4\pi}{N_{r_i}}\frac{4\pi}{N_{r_j}}\sum_{lm} \left[\boldsymbol{\Omega}_l^{(\pm)-1}\right]_{r_i,r_j}{}_{\pm1}Y_{lm}(\hat{\mathbf{n}}_i){}_{1}Y_{lm}^*(\hat{\mathbf{n}}_j),
  \label{eq:cplusninv}
\end{equation}
where we have assumed that the sum over angular indices can be converted to an integral, which allows use of the orthogonality relation of the spin-weighted spherical harmonics.

Using the result of Equation~\eqref{eq:cplusninv} we can form the blocks of the full inverse covariance matrix and hence derive the $\mathbfss{M}$ matrices needed for inferring the SP and SAD proper motion. We find (upon use of the Woodbury matrix formula)
\begin{align}
  \mathsf{M}^{(0)}_{ab} &= \delta_{ab} \frac{8\pi}{3} \sum_{r_i,r_j} \left(\mathbfss{S}^{E,u}_1 + \mathbfss{N}^E\right)^{-1}_{r_i,r_j}, \nonumber \\
  \mathsf{M}^{(1)}_{ab} &= \delta_{ab} \frac{8\pi}{3} \sum_{r_i,r_j} \frac{\left(\mathbfss{S}^{E,u}_1 + \mathbfss{N}^E\right)^{-1}_{r_i,r_j}}{r_j}, \nonumber \\
  \mathsf{M}^{(2)}_{ab} &= \delta_{ab} \frac{8\pi}{3} \sum_{r_i,r_j} \frac{\left(\mathbfss{S}^{E,u}_1 + \mathbfss{N}^E\right)^{-1}_{r_i,r_j}}{r_ir_j},
  \label{eq:LNINMs}
\end{align}
where $\mathbfss{N}^E$ is the E-mode noise power, whose $ij$ component is equal to $\delta_{r_i r_j} 4\pi\sigma_\mu^2(r_i)/N_{r_i}$, and we have packaged up the E-mode LSS power spectrum into a $N_r \times N_r$ matrix $\mathbfss{S}_l^E$. Importantly, \emph{only the dipole of LSS transverse velocities contributes to the variance of the SP and SAD proper motions}. This result makes intuitive sense - higher multipoles of the LSS transverse velocity field cannot be confused for the dipolar SP and SAD proper motions, and are orthogonal in the limit of large $N$ due to isotropy, which is precisely the LNIN limit.

The above construction is formally only valid in the situation where the noise variance is constant in each radial bin. In reality this is not the case, since there is a distribution of magnitudes and colours across the sky within each bin, which gives $\sigma_v^2$ angular dependence. Given this situation, what is the correct `averaged' noise variance which should be used for $\mathbfss{N}^E$? Conditioning on the noise variances and galaxy positions was the approach adopted in the previous sections, but to simplify the inverse covariance matrix we had to artificially fix the variances such that the spin-weighted spherical harmonics carried all the angular dependence. Formally the correct thing to do would be to marginalise over the magnitudes and colours in each radial bin, but this would give a non-Gaussian likelihood due to the non-linear dependence of noise variance,  magnitude, and colour, as well as the intrinsically non-Gaussian underlying distributions of these quantities. To simplify the situation we simply take the limit of Equation~\eqref{eq:LNINMs} in which the cosmic variance is zero. In this limit, Equation~\eqref{eq:LNINMs} reduces to Equation~\eqref{eq:LNINMnocv} only if we use the \emph{average of the inverse noise variance over the radial bin}. We henceforth adopt this choice, but investigate the consequences of making a different choice for the average noise variance in Appendix~\ref{app:nvar}.

We can also find the LNIN form of the $\mathbf{L}$ vectors. Expanding the data vector in spin-weighted spherical harmonics, we find that
\begin{equation}
  \mathbf{L}^{(n)} = \sqrt{\frac{8\pi}{3}}  \sum_{r_i,r_j} \frac{\left(\mathbfss{S}^{E,u}_1 + \mathbfss{N}^E\right)^{-1}_{r_i,r_j}}{r_i^n} \mathbfss{B}^{-1}\mathbf{d}^E_m(r_j),
  \label{eq:LNINLs}
\end{equation}
where $\mathbf{d}^E_m = (d^E_{1-1}, d^E_{10}, d^E_{11})^\intercal$ is a vector of the dipolar E-mode multipole coefficients of the data, and $\mathbfss{B}$ is the unitary matrix which converts from dipolar multipole space ($m=-1,0,1$) to three-dimensional Cartesian space (see the discussion in Section~\ref{subsec:dipoles}). Equation~\eqref{eq:LNINLs} demonstrates that the $\mathbf{L}$ vectors are simply the E-mode dipolar coefficients of the data `rotated' to real Cartesian space, inverse-variance weighted, distance weighted (depending on whether SAD or SP is being measured), and then averaged over all radial bins, which makes intuitive sense for Gaussian data. These vectors may then be further normalized and projected against known velocity or proper motion vectors to estimate the SP or SAD amplitudes.

We note finally that the bias in the estimators due to correlated LSS noise may also be derived in the LNIN approximation, which offers some insights. For example, we find that the bias in $A_{\mathrm{SP}}$ is proportional to $\cos\psi$, where $\psi$ is the angle between the Solar System-CMB relative velocity and the Milky-Way-CMB relative velocity. We find that $\psi \approx 20^\circ$. In the case of $A_{\mathrm{SAD}}$ we find $\psi \approx 87^\circ$, which implies that the bias in the SAD proper motion from correlated LSS proper motions is strongly suppressed by the fortuitous perpendicular alignment of the acceleration vector and the Milky Way's velocity vector relative to the CMB. Explicitly, the bias to the $\mathbf{L}$ vectors is
\begin{equation}
  \Delta \langle \mathbf{L}^{(n)} \rangle = \frac{8\pi}{3} \mathbf{v}^{\mathrm{SP}}_c  \sum_{r_i,r_j} \frac{\left(\mathbfss{S}^{E,u}_1 + \mathbfss{N}^E\right)^{-1}_{r_i,r_j}}{r_i^nr_j} \frac{-3\Psi_\perp(r_j)}{\sigma_v^2}.
\end{equation}

The LNIN approximations to the estimators can be computed rapidly, since they only involve averages over the radial distribution of galaxies and the inversion of low-dimensional matrices. The price paid for this is a restrictive assumption about the form of the noise variances. Where possible, these assumptions can be tested against the exact results presented in Section~\eqref{subsec:estimators}. We find that for the survey assumptions made in Section~\eqref{sec:gums}, the LNIN approximations predict variances, biases, and correlations to better than 10\% for the cosmic-variance-free estimator of Equation~\eqref{eq:Lnoise} (for which the covariance matrix inversions can be done analytically). This is due to the large number of galaxies typically included in each radial bin, the smoothness of the signal with distance making the radial binning close to lossless, the instrument noise being close to isotropic, and the sky coverage being almost complete.

\section{Survey assumptions}
\label{sec:gums}

In order to forecast the ability of astrometric surveys to detect large-scale extragalactic proper motions, we must make assumptions for the radial distributions of galaxies and quasars, as well as their proper motion noise variances. In order to test the LNIN approximations of Section~\ref{subsec:LNIN}, we also need a realistic angular distribution. Following~\citet{2016A&A...589A..71B}, as our reference astrometric survey we use the Gaia Universe Model Snapshot (GUMS)~\citep{2012A&A...543A.100R}, a simulated catalogue which includes galaxies (unresolved in their stars) and quasars with realistic spatial, magnitude, and colour distributions similar to those observable with Gaia. The catalogue contains roughly 1 million quasars and about 38 million galaxies, although most of these are unobservable due to their extended surface brightness profiles.

We select galaxies and quasars from the catalogue with $0 \leq z \leq 0.1$ but vary the distance ranges used in the analysis in order to gain insights into the signal-to-noise forecasts\footnote{Note that we do not account for survey incompleteness in our forecasts. The catalogue will likely be incomplete for the faint objects around $G=20$. This is not as serious an issue for proper motions as it is for radial velocities since we do not rely on distance proxies which correlate with flux, but incompleteness which varies over the sky will impact the optimality of our estimators and should be accounted for in a more thorough treatment. Isotropic incompleteness will affect the clustering properties of measured sources, but this should provide a only small correction when correlating proper motions over large scales. We defer further investigation into these issues to a future work, and instead assume full completeness down to $G=20$, as in~\citet{2012ApJ...755...58N}.}. The SP and LSS proper motions both fall as $1/r$, but more distant objects can help break degeneracies with the SAD proper motion and suffer less from bias due to correlations of LSS peculiar velocities with our local motion. Gaia is only expected to observe galaxies with centrally concentrated brightness profiles, and is unlikely to be sensitive to galaxies with a large disk component~\citep{2014A&A...568A.124D}. In particular, late-type spiral galaxies are unlikely to be present in Gaia data~\citep{2015A&A...576A..74D}. As in~\citet{2016A&A...589A..71B}, we select galaxies with Hubble type E2, E-S0, Sa and Sb. We add to these galaxies the small number of quasars which reside at $z \leq 0.1$ in the GUMS catalogue, which results in a final catalogue of roughly $2 \times 10^6$ objects, roughly half of which are Sa or Sb spiral galaxies. The inclusion of Sa and Sb spirals in the catalogue is optimistic despite their significant bulge components (bulge-to-total flux ratios of roughly 0.5 and 0.3 respectively,~\citealt{2015A&A...576A..74D}), but since our aim in this work is primarily to test our correlation and likelihood formalism and study the optimal redshift ranges for measuring $H_0$, this assumption is not critical and allows for comparison with~\citet{2016A&A...589A..71B}. Furthermore, cosmic variance from LSS transverse velocities is the dominant source of variance in our estimators, so we expect our final results to be fairly insensitive to the total number of galaxies in a fixed radial distribution.

\begin{figure}
  \includegraphics[width=\columnwidth]{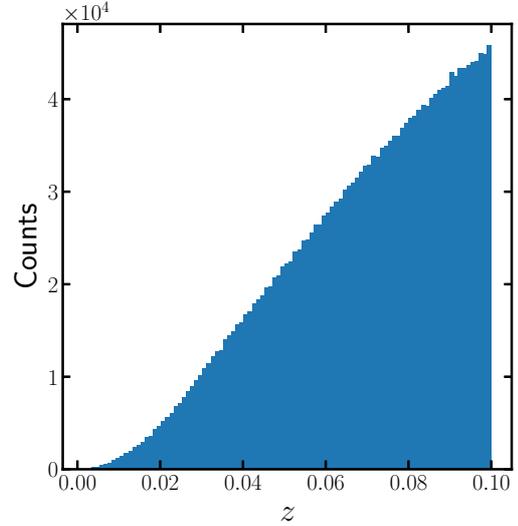}
  \caption{Redshift distribution of the galaxies used in our forecasts, taken from GUMS (see text for details). We cut out galaxies having $z > 0.1$, as there is little SP signal as these distances.}
  \label{fig:gums_gals_zdist}
\end{figure}

\begin{figure}
  \includegraphics[width=\columnwidth]{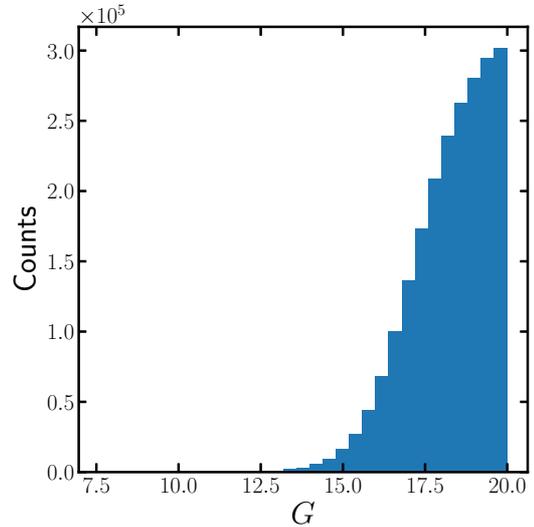}
  \caption{Gaia $G$-band magnitude distribution of the galaxies used in our forecasts, taken from GUMS (see text for details). The magnitude limit for Gaia is roughly $G=20$.}
  \label{fig:gums_gals_Gdist}
\end{figure}

In Figure~\ref{fig:gums_gals_zdist} and~\ref{fig:gums_gals_Gdist} we plot the redshift distribution and Gaia $G$-band magnitude distribution of the galaxies used in our forecasts. Most of the galaxies in our sample reside close to the redshift limit of the survey, but since the proper motion signals we are measuring fall as $1/r$ this upper limit is sufficient, as we shall see. Likewise, most of the galaxies have magnitudes close to the Gaia magnitude limit $G=20$.

In order to assign proper motion variances to each object in the catalogue, we assume that the Gaia end-of-mission five-year point-source specifications may be applied to galaxies (as in~\citealt{2012ApJ...755...58N} and~\citealt{2016A&A...589A..71B}). The accuracy of the proper motion measurements of extended objects made with Gaia is uncertain, and the use of point-source specifications is likely over-optimistic. For a circular Gaussian image with noise limited by photon counts (rather than the background), the error of the centroid is proportional to the angular diameter of the image (i.e. the only length scale in the problem, see e.g.~\citealt{1985MNRAS.214..575I}). For rectilinear proper motion, it is straightforward to show that the proper motion in a particular direction is also proportional to the size, with an extra factor $t^{-1}$ where $t$ is the total observation time (c.f. the fact that proper motion errors scale as $t^{-3/2}$ whereas centroid errors scale as $t^{-1/2}$). For galaxies with a fixed physical diameter of 1 kpc, the smallest angular diameter in our sample is roughly 0.5 arcsecs, which occurs at the upper redshift limit of $z=0.1$ (the minimum value at any redshift for our cosmology is 0.11 arcsecs, which occurs at $z\approx 1.6$). The Gaia resolution is roughly 0.1 arcsecs (FWHM), so all the galaxies in our sample would appear as resolved objects if they had a fixed physical FWHM of 1 kpc. Assuming a circular PSF, the image size of the smallest galaxy in our sample is five times larger than that of a point-source, which for fixed apparent magnitude and integration time suggests the proper motion error could be at least five times larger for most of the galaxies in our sample. Of course, galaxies with sub-kpc bulges could well appear as point-sources with Gaia;~\citet{2012ApJ...755...58N} argue that the vast majority of early-type galaxies and most late-type galaxies will be observed as point sources with Gaia\footnote{In principle this information is already present in the Gaia DR2 catalogue.}. Additionally, galaxies with sharp features in their surface brightness profile will have inherently more precise astrometry; many of the low-redshift galaxies observed with Gaia will be well-resolved, and resolved structure in the image will improve the centroid determination.

Since our primary aim here is to study the broad behaviour of proper motion inference in the presence of LSS and distance cuts rather than sensitivity to the precise details of the measurement pipeline, the assumption of point-source noise is justifiable. However, we caution the reader against the overinterpretation of our final forecasts; a more complete treatment would involve running the GUMS galaxies through a realistic Gaia detection pipeline, but this is beyond the scope of this work. Furthermore, for many of the applications of our catalogue the instrument noise is sub-dominant to the cosmic variance, and its underestimation will have only a small impact on results.

We also neglect the small spatial correlations expected between proper motion errors in the final end-of-mission Gaia data~\citep{2010IAUS..261..320H}, although these could easily be included in our likelihood formalism (spatial correlations in proper motion errors are present in Gaia DR2~\citealt{2018A&A...616A...2L}).

\begin{figure}
  \includegraphics[width=\columnwidth]{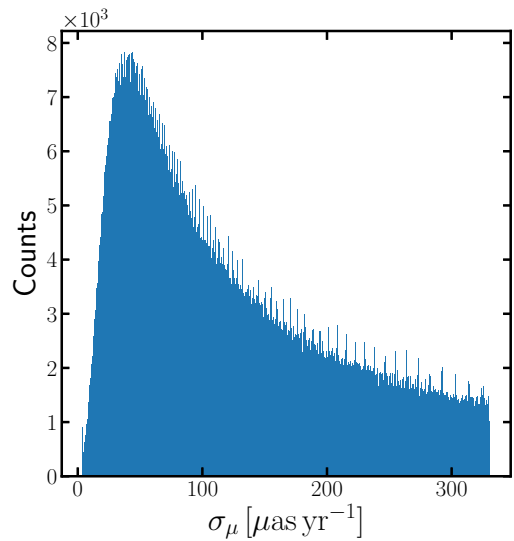}
  \caption{Distribution of the mean proper motion error for galaxies used in our forecasts (see text for details). These numbers were computed using the Gaia end-of-mission five-year point-source specifications and the magnitudes and colours produced by the GUMS galaxy catalogue. The sharp cutoff at large error values is due to the magnitude limit of the survey, while the small spike at $\sigma_{\mu} \approx 4 \, \mu \mathrm{as} \, \mathrm{yr}^{-1}$ is due to the bright-star parallax noise floor. The remaining spikes are due to the choice of binning.}
  \label{fig:gums_gals_pmerrdist}
\end{figure}

We use the formulae from the Gaia science performance page\footnote{\url{https://www.cosmos.esa.int/web/gaia/science-performance}} to convert each galaxy's $G$-band magnitude and colour to a proper motion uncertainty. We consistently distinguish between uncertainty in the declination and right-ascension proper motion, except when adopting the LNIN approximations (the component uncertainties differ by roughly 10\%). In Figure~\ref{fig:gums_gals_pmerrdist} we plot the distribution of mean proper motion errors from the GUMS galaxies. The distribution is truncated at roughly $330 \, \mu\mathrm{as} \, \mathrm{yr}^{-1}$ corresponding to the magnitude limit of the survey, and has a broad peak at roughly $50 \, \mu\mathrm{as} \, \mathrm{yr}^{-1}$. Barely visible in Figure~\ref{fig:gums_gals_pmerrdist} is a small spike at $\sigma_{\mu} \approx 4 \, \mu \mathrm{as} \, \mathrm{yr}^{-1}$, the minimum proper motion uncertainty, which is due to the bright-star parallax noise floor. We remind the reader that the amplitude of the SAD proper motion is roughly $4.3 \, \mu\mathrm{as} \, \mathrm{yr}^{-1}$ and correlated over large-scales, while the SP proper motion is roughly $\frac{78}{r / 1 \, \mathrm{Mpc}} \, \mu\mathrm{as} \, \mathrm{yr}^{-1}$. These estimates suggest that both SP and SAD proper motions could be measurable in our galaxy sample and assumed proper motion variances.

In order to convert proper motions into transverse velocities, we assume that spectroscopic redshifts are available for every galaxy in our catalogue. This is a somewhat optimistic assumption given that we include all galaxies down to $G=20$. Gaia will provide medium-resolution spectroscopy for objects down to $G=17$~\citep{2014A&A...568A.124D}, which constitutes roughly 15\% of our total sample. We will therefore have to rely on existing spectroscopic samples for the remaining objects - for example SDSS should provide substantial overlap in the north due its greater depth, and the upcoming DESI survey will improve on this further. We have not quantified the likely overlap with existing and planned surveys, but argue that this work provides further motivation for obtaining accurate redshifts with wide spectroscopic galaxy surveys. It should also be noted that since cosmic variance rather than instrumental noise limits the Hubble constant measurement, our forecasts for $H_0$ should be fairly robust to a modest reduction in the galaxy sample size. Measurement of the transverse velocity of LSS will be more affected by incomplete redshift coverage, so we again caution the reader against overinterpretation of these forecasts.
  
Having established our galaxy sample, we now use the expressions derived in~\ref{sec:like} to forecast the detection significance of the SP and SAD proper motions (the former proportional to the $H_0$), properly accounting for bias and variance from LSS transverse velocities. While the reference survey is a Gaia-like astrometric survey, the formalism we have presented can be applied to any proper motion measurements. For example, more precise astrometry is expected from the ngVLA~\citep{2018arXiv180706670D} and post-Gaia optical and near-infrared missions~\citep{2016arXiv160907325H}.

\section{Signal-to-noise forecasts}
\label{sec:results}

Using the catalogue of galaxies and quasars presented in Section~\ref{sec:gums}, we compute the variances and biases of the SP proper motion under a range of varying assumptions. We consider the sensitivity of the forecasts to varying the maximum redshift in the catalogue $z_{\mathrm{max}}$, the minimum distance $r_{\mathrm{min}}$, and to the marginalization or fixing of the SAD proper motion. The SAD signal is independent of distance and so could be well-constrained by a sample of high-redshift quasars, so fixing it is a reasonable approach for this analysis.

We also compute the extra variance from LSS proper motions when using the cosmic-variance-free proper motion estimator of Equation~\eqref{eq:Lnoise} (labelled as `Including LSS' in the figures), and the variance when marginalizing over these proper motions (labelled as `Marginalized LSS' in the figures), and study the impact of fixing the direction of the SP proper motion to that inferred from the CMB dipole~\citep{2014A&A...571A..27P}. We also compute the bias to the SP proper motion from LSS transverse velocities correlated with our local motion, as described in Section~\ref{subsec:bias}. As mentioned in Section~\ref{subsec:LNIN}, the LNIN approximations were found to be good to better than 10\% for all forecasts made with the cosmic-variance-free estimator, so we adopt the LNIN approximation throughout this section. This allows rapid calculation of the inverse covariance matrices needed to form of the optimal estimators presented in Section~\ref{sec:like}.

\subsection{Varying $z_{\mathrm{max}}$}
\label{subsec:zhi}

In this section we study the impact on our forecasts of varying the maximum redshift of the catalogue, $z_{\mathrm{max}}$, fixing the minimum distance to be $r_{\mathrm{min}} = 20 \, \mathrm{Mpc}/h$.

\begin{figure}
  \includegraphics[width=\columnwidth]{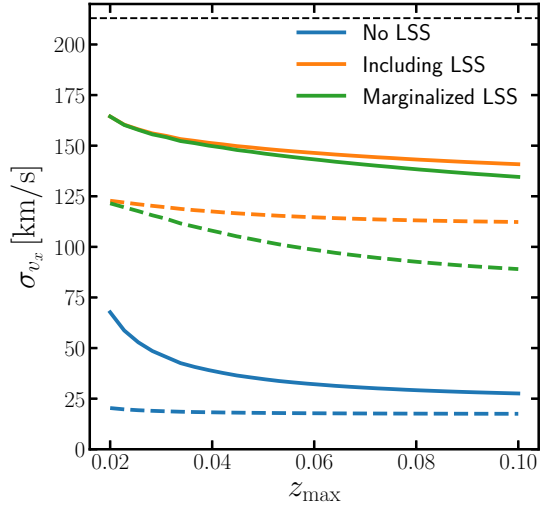}
  \caption{Forecast error on one component of the SP velocity as a function of the highest redshift included in the survey, for $r_{\mathrm{min}} = 20 \, \mathrm{Mpc}/h$. The curves are forecasts setting LSS to zero (blue, lowest curve), including LSS without marginalization (orange, highest curve), and with marginalization over LSS (green, middle curve). Solid and dashed curves are with and without marginalization over the SAD proper motion respectively. The black dashed horizontal line shows the r.m.s. value from the CMB dipole.}
  \label{fig:SPSD_zhi}
\end{figure}

In Figure~\ref{fig:SPSD_zhi} we plot the forecast error (square-root of the estimator variance) on one component of the SP velocity (equal to the errors in the other components in the LNIN approximation). We plot the error assuming no cosmic variance (blue curves), the error including cosmic variance from LSS proper motions uncorrelated with our local velocity (orange curves), and the error after marginalizing over these LSS proper motions (green curves). Solid and dashed curves are with and without marginalization over the SAD proper motion, and we have assumed $H_0$ is fixed such that spectroscopic redshifts directly give the distance (accounting for the radial velocity uncertainty in the broad binning, see the footnote in Section~\ref{subsec:LNIN}). We also plot the r.m.s. value of the SP velocity as inferred from the CMB dipole (black dashed line).

As expected, all the errors decrease as the maximum redshift is raised since more galaxies are included in the estimators, but the improvement is only marginal since the signal falls as $1/r$. The variance from LSS dominates over that from instrumental noise, and can only be mildly reduced by marginalization. With LSS marginalization and $z_{\mathrm{max}} = 0.1$ we find $\sigma_{v_x} \approx 130 \, \mathrm{km}/s$ for marginalized SAD and $\sigma_{v_x} \approx 85 \, \mathrm{km}/s$ for fixed SAD, compared to the typical signal amplitude of $213 \, \mathrm{km}/s$, which represent $1.6\sigma$ (marginalized SAD) and $2.5\sigma$ (fixed SAD) measurements. With only variance from instrumental noise this would be boosted to roughly $8\sigma$, which shows the importance of including peculiar velocities from LSS in these forecasts. Our forecast errors are similar to (albeit larger than) those of~\citet{2016A&A...589A..71B}, who found $\sigma_{v_x} \approx 100 \, \mathrm{km}/s$ with $z_{\mathrm{max}} \approx 0.03$ and simultaneous fitting of SAD (equivalent to our solid curves). This difference could be explained by the neglect of correlations between LSS proper motions in~\citet{2016A&A...589A..71B}.

Most of the signal-to-noise has converged by $z_{\mathrm{max}} \approx 0.04$, with mild dependence on whether or not SAD is marginalized over. Marginalizing over SAD increases the error in all cases (as expected), with constraints degrading by a factor of roughly 50\% for $z_{\mathrm{max}} = 0.1$. When $z_{\mathrm{max}}$ is small this degradation is significantly larger for the cosmic-variance-free estimator, as a smaller range of distances is available to break degeneracies with SP. When including LSS variance the degradation has a much more mild dependence on $z_{\mathrm{max}}$.

\begin{figure}
  \includegraphics[width=\columnwidth]{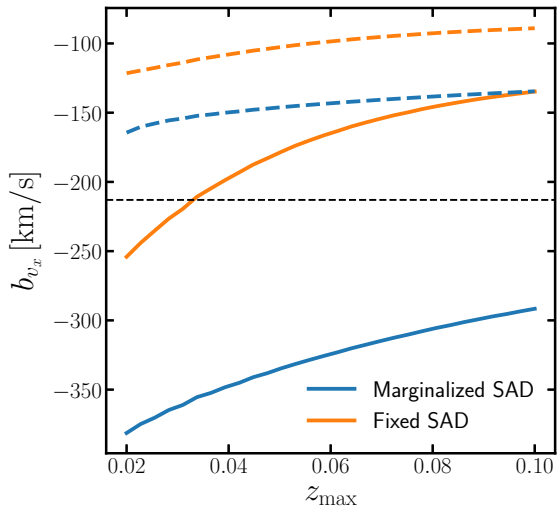}
  \caption{The bias of a single component of the SP velocity from LSS proper motions correlated with the Solar System's local motion (see Section~\ref{subsec:bias} for a discussion). We plot biases for the estimator with uncorrelated LSS marginalized out, and for both fixed SAD (orange, upper solid) and marginalized SAD (blue, lower solid). We also plot the (negative) error of this estimator (dashed curves). For all values of $z_{\mathrm{max}}$ considered, the bias in the estimator exceeds its forecast error. The horizontal dashed line shows the r.m.s. SP velocity from the CMB dipole.}
  \label{fig:bSP_zhi}
\end{figure}

The optimal estimator is that with uncorrelated LSS transverse velocities marginalized out. In Figure~\ref{fig:bSP_zhi} we plot the bias from correlated LSS velocities (fixing or marginalizing over SAD) as a function of $z_{\mathrm{max}}$, for a single component of the SP velocity estimator\footnote{The bias is proportional to the corresponding component of the Milky Way-CMB relative velocity; for our Cartesian coordinate system the $x$-component carries roughly 75\% of the squared-length of this velocity vector. The bias in the $y$ and $z$ directions are roughly a factor of $-0.4$ and $0.5$ smaller respectively. Note that the sign of the bias depends on the chosen direction.}. For all values of $z_{\mathrm{max}}$ considered, the bias in the estimator exceeds its forecast error, and in some cases the bias can be more than twice the $1\sigma$ error. This bias originates from the inclusion of galaxies at low distances whose LSS proper motions are strongly correlated with that of the Milky Way (c.f. Figure~\ref{fig:gorski_perp_normed}), but tends to decrease as more high-redshift objects (which individually have low bias) are included in the catalogue due to the increasing $\mathbfss{M}$ normalization of the $\mathbf{L}$ vector. This also explains why the bias is greater for the marginalized-SAD estimator, as this has a greater weighting due to the smaller effective $\mathbfss{M}$ which results from projecting out the SAD proper motion. Figure~\ref{fig:bSP_zhi} suggests that for this choice of $r_{\mathrm{min}}$, the bias from correlated LSS will have to be mitigated if the potential SP detection suggested in Figure~\ref{fig:SPSD_zhi} can be realised.
%

\begin{figure}
  \includegraphics[width=\columnwidth]{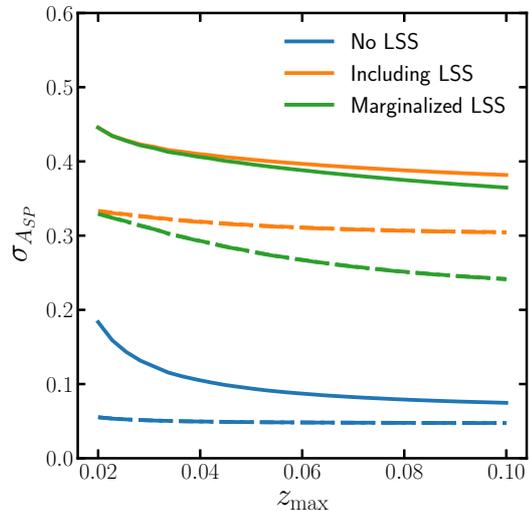}
  \caption{Forecast error on the amplitude of the SP velocity assuming the direction is fixed to that measured by the CMB dipole. The curves are forecasts setting LSS to zero (blue, lowest curves), including LSS without marginalization (orange, highest curves), and with marginalization over LSS (green, middle curves). Solid curves have had both the amplitude and direction of the SAD proper motion marginalized over, dashed curves have had only the amplitude of the SAD proper motion marginalized over with the direction kept fixed, and dot-dashed curves (indistinguishable from the dashed curves) have had both the amplitude and direction of the SAD proper motion kept fixed. The error on $A_{\mathrm{SP}}$ can be interpreted as the forecast fractional error on $H_0$.}
  \label{fig:ASPSD_zhi}
\end{figure}

In Figure~\ref{fig:ASPSD_zhi} we plot the forecast error on the amplitude of the SP velocity, with the direction fixed to that of the CMB dipole. Fixing the amplitude this can be reinterpreted as the forecast fractional constraint on $H_0$, as described in Section~\ref{subsec:asp}. The behaviour in this plot is similar to that in Figure~\ref{fig:SPSD_zhi}, with a greater detection significance upon fixing the SP direction. Fixing the SAD proper motion (dot-dashed curves) yields a detection significance of roughly $4 \sigma$ (i.e. a 25\% measurement of $H_0$), whereas marginalizing over it (solid curves) gives roughly $2.7 \sigma$ (a 40\% measurement of $H_0$).  Fixing the SAD direction to the direction of the galactic centre gives almost identical results to fixing both amplitude and direction, suggesting most of the uncertainty from SAD comes from constraining its direction rather than its amplitude. This is due to the fact that the Solar System's acceleration vector and the CMB dipole are roughly orthogonal, subtending an angle of about $94^\circ$, which results in no confusion between the SP and SAD signals when their directions are fixed. Once again our forecast error on $H_0$ is larger than that of~\citet{2016A&A...589A..71B}, who forecast an error of roughly 30\% on $H_0$ for $z_{\mathrm{max}} = 0.03$, compared to our roughly 40\% prediction. This discrepancy is likely a result of our inclusion of correlated velocities due to LSS.

%
%
\begin{figure}
  \includegraphics[width=\columnwidth]{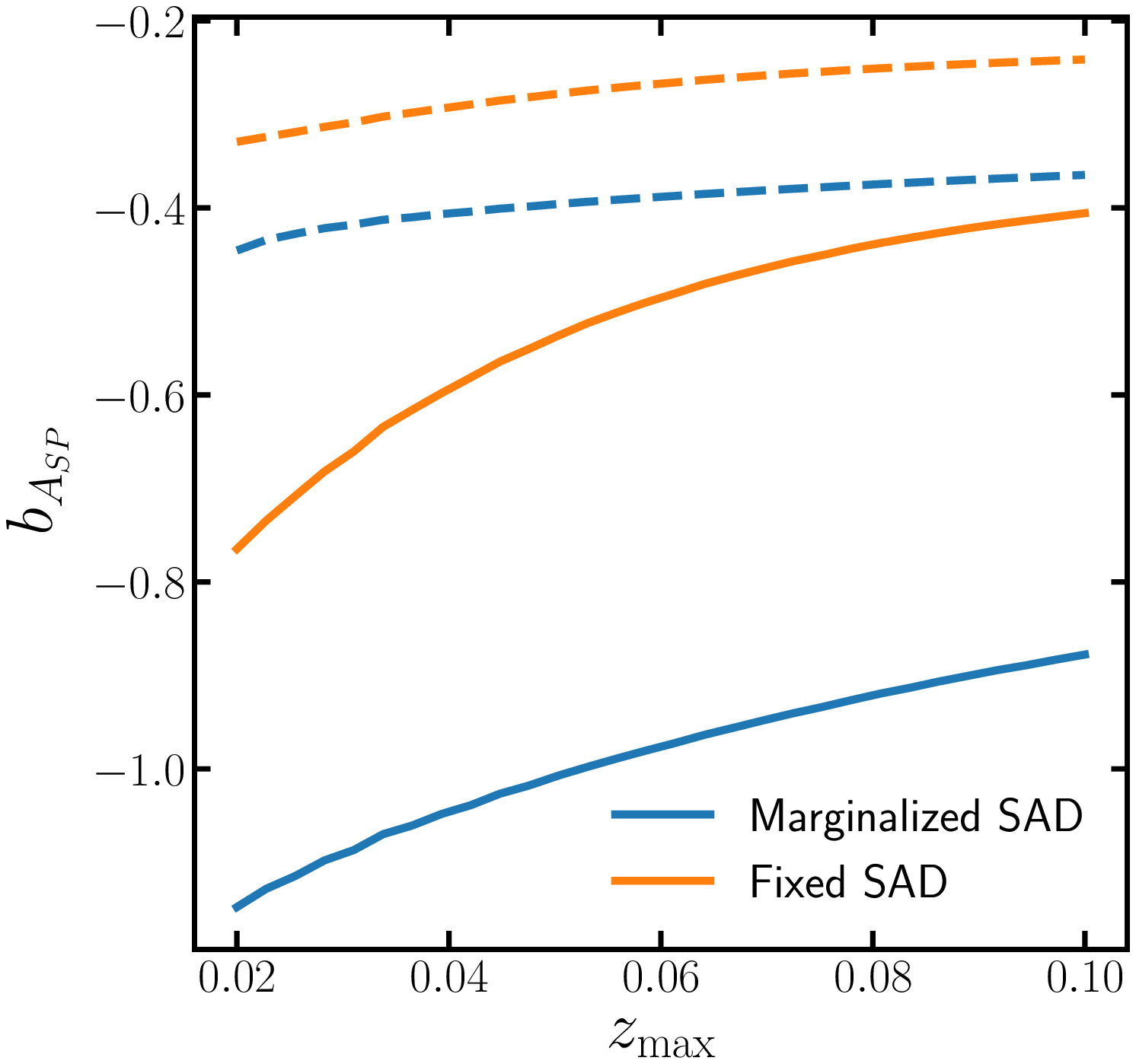}
  \caption{The bias of the amplitude of the SP velocity assuming the direction is fixed to that measured by the CMB dipole from LSS proper motions correlated with the Solar System's local motion (see Section~\ref{subsec:bias} for a discussion). We plot biases for the estimator with uncorrelated LSS marginalized out, and for both fixed SAD (orange, upper solid) and marginalized SAD (blue, lower solid). We also plot the (negative) error of this estimator (dashed curves). For all values of $z_{\mathrm{max}}$ considered, the bias in the estimator exceeds its forecast error.}
  \label{fig:bASP_zhi}
\end{figure}

In Figure~\ref{fig:bASP_zhi} we plot the bias to $A_{\mathrm{SP}}$ from correlated LSS transverse velocities. The shape of these curves is similar to those of the SP velocity components in Figure~\ref{fig:bSP_zhi}, with the sign of the bias negative due to the angle between the total SP velocity and the part correlated between LSS being less than $90^\circ$ and the negative sign in Equation~\eqref{eq:muclss}. Once again we see that if unaccounted for, the bias from LSS transverse velocities correlated to our local motion would overwhelm the forecast error, by up to a factor of two, when $r_{\mathrm{min}} = 20 \, \mathrm{Mpc}/h$.

\subsection{Varying $r_{\mathrm{min}}$}
\label{subsec:rmin}

In this section we study the impact on our forecasts of varying the minimum comoving distance of the catalogue, $r_{\mathrm{min}}$, fixing the maximum redshift to be $z_{\mathrm{max}} = 0.1$. We focus on only $A_{\mathrm{SP}}$; the qualitative conclusions are very similar for the individual components of the SP velocity vector.

\begin{figure}
  \includegraphics[width=\columnwidth]{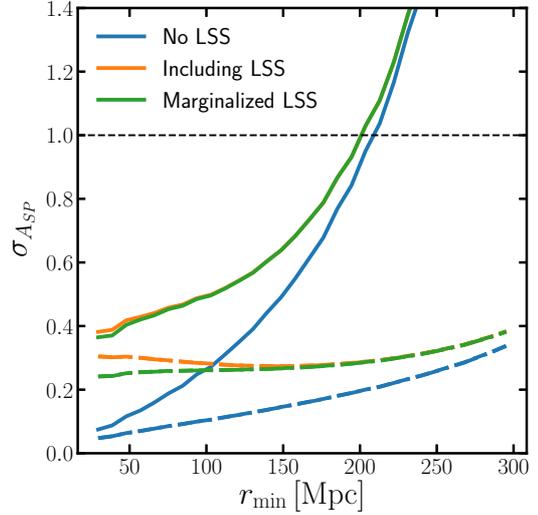}
  \caption{Same as Figure~\ref{fig:ASPSD_zhi} but varying $r_{\mathrm{min}}$. The curves are forecasts setting LSS to zero (blue, lowest curve), including LSS without marginalization (orange, highest curve), and with marginalization over LSS (green, middle curve). Solid curves have had both the amplitude and direction of the SAD proper motion marginalized over, dashed curves have had only the amplitude of the SAD proper motion marginalized over with the direction kept fixed, and dot-dashed curves (indistinguishable from the dashed curves) have had both the amplitude and direction of the SAD proper motion kept fixed. The dashed horizontal line denotes the fiducial value of $A_{\mathrm{SP}} = 1$. The error on $A_{\mathrm{SP}}$ can be interpreted as the forecast fractional error on $H_0$.}
  \label{fig:ASPSD_rlo}
\end{figure}

In Figure~\ref{fig:ASPSD_rlo} we plot the forecast errors assuming the SP direction is fixed by the CMB dipole, similarly to Figure~\ref{fig:ASPSD_zhi}.  As expected, as we throw away galaxies for which the signal (proportional to $1/r$) is high, the variance increases. Steep increases are seen when the SAD proper motion is projected out (solid curves), since in this scenario distance information is essential to break degeneracies. Significant detections of $A_{\mathrm{SP}}$, or $H_0$ if the amplitude is fixed from the CMB dipole, are possible, although we require $r_{\mathrm{min}} \lesssim 200 \, \mathrm{Mpc}$ if marginalizing over SAD to get a $1\sigma$ measurement.

When fixing SAD and increasing $r_{\mathrm{min}}$ from its lowest value, we find that the total variance in the cosmic-variance-free estimator (orange dashed curve) \emph{decreases} as the dipole of the uncorrelated LSS transverse velocity decreases due to projection (see Section~\ref{subsec:LSSspec}) before eventually increasing due to the loss of high signal-to-noise galaxies. This behaviour is not present in the cosmic-variance-marginalized estimator (green dashed curve), which correctly accounts for this extra variance by including it in the inverse covariance matrix.
%
%
%

\begin{figure}
  \includegraphics[width=\columnwidth]{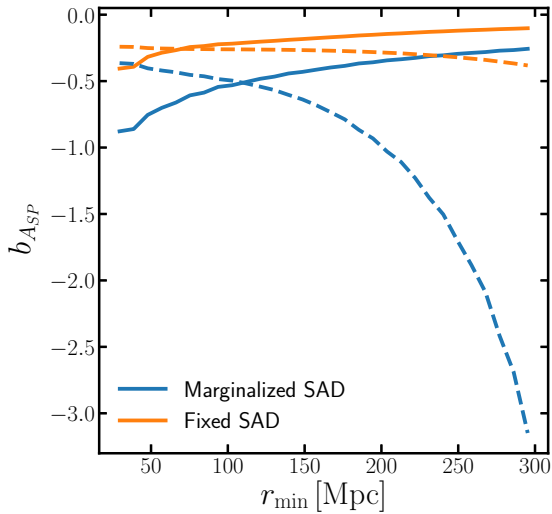}
  \caption{The bias of the amplitude of the SP velocity assuming the direction is fixed to that measured by the CMB dipole from LSS proper motions correlated with the Solar System's local motion (see Section~\ref{subsec:bias} for a discussion). We plot biases for the estimator with uncorrelated LSS marginalized out, and for both fixed SAD (orange, upper solid) and marginalized SAD (blue, lower solid). We also plot the (negative) error of this estimator (dashed curves).}
  \label{fig:bASP_rlo}
\end{figure}

In Figure~\ref{fig:bASP_rlo} we plot the bias in $A_{\mathrm{SP}}$ from correlated LSS transverse velocities. The lowest redshift galaxies contribute most to this bias, as these are most correlated with the Solar System's velocity (Figure~\ref{fig:gorski_perp_normed}). Omitting these galaxies from the analysis reduces the bias, bringing it below the $1\sigma$ error for $r_{\mathrm{min}} \gtrsim 100 \, \mathrm{Mpc}$ for marginalized SAD and $r_{\mathrm{min}} \gtrsim 50 \, \mathrm{Mpc}$ for fixed SAD. Restricting to values of $r_{\mathrm{min}}$ for which this bias is less than the error, we forecast that $A_{\mathrm{SP}}$ can be measured at roughly $2\sigma$ for marginalized SAD and $4\sigma$ for fixed SAD, corresponding to 50\% and 25\% measurements of the Hubble constant respectively. This bias could be reduced by modelling it (see Section~\ref{subsec:bias}) or by attempting to constrain the amplitude of the SP component orthogonal to the part correlated with LSS. However we find that this latter approach entails the loss of roughly 50\% of the signal-to-noise (which follows from the angles subtended by the relevant velocity vectors).

In the case of the individual SP velocity components we reach similar conclusions, with galaxies below $100 \, \mathrm{Mpc}$ and $50 \, \mathrm{Mpc}$ requiring omission if SAD is marginalized or fixed respectively. This limits us to a $1$-$2\sigma$ measurement of the SP velocity in a single component. Thus, cutting out low-redshift galaxies from the catalogue can bring the bias down to acceptable levels, but a significant detection of the SP velocity requires prior knowledge of its direction from the CMB dipole.

%

Thus it appears that cutting out galaxies to mitigate the bias due to correlated LSS transverse velocities limits us to at best a 25\% measurement of $H_0$ with our survey specifications. For fixed SAD, knowing the bias helps little, since the saturation of the error at low values of $r_{\mathrm{min}}$ limits the gains of reducing the minimum distance (as does the uncertainty in distances due to radial velocities in the redshift). It thus seems that competitive measurements of of $H_0$ will require going to redshifts much higher than $z=0.1$ in order to reduce cosmic variance (see Figure~\ref{fig:SPSD_zhi}). Alternatively, combining with radial velocity measurements (with redshift-independent distance measures) could potentially reduce the cosmic variance, since both components trace the same underlying dark matter. We caution against overinterpretation of these results however since several simplifying assumptions have been made, in particular the assumptions which go into the galaxy distributions and noise variances (see Section~\ref{sec:gums}). Nevertheless, our analysis has highlighted the importance of cosmic variance in measurements of $H_0$ with the proper motions of nearby galaxies.

\section{Measuring the proper motion of large-scale structure}
\label{sec:LSS}

So far we have focussed on the possibility of measuring dipolar patterns in the proper motion of extragalactic objects, identifying sources of bias and variance which make this measurement difficult. Alternatively, as in~\citet{2012ApJ...755...58N, 2018ApJ...864...37D,2018arXiv180807103T},  we could try to measure anisotropies in the proper motion on smaller angular scales. The $l\geq2$ modes of the proper motion field are sourced entirely by the transverse velocities of galaxies and quasars induced by LSS, and do not suffer from contamination from local effects such as SP or SAD. Figure~\ref{fig:vperp_ells_xiE} suggests all modes with $l\lesssim 6$ will be signal-dominated for our lowest redshift bin, which raises hopes that a measurement of LSS transverse velocities might be possible with our catalogue. With spectroscopic redshift information, such a measurement would be sensitive to the amplitude of the proper motion correlation function, specifically the parameter combination $H_0 f \sigma_8$. In this section we forecast the ability of our galaxy catalogue to measure the higher-order $l$-modes of the proper motion.

As a first step, we will simply attempt to constrain the amplitude of the part of the LSS transverse velocity uncorrelated with our local motion. Recall that this correlation only affects the $l=1$ mode, so this procedure retains the amplitude information in the higher-order modes. We parametrize the uncorrelated LSS covariance matrix as
\begin{equation}
  \mathbfss{C}^u_\pm = A_{\mathrm{LSS}}^2 \mathbfss{C}^u_{\pm,0},
\end{equation}
where $\mathbfss{C}^u_{\pm,0}$ is a fixed fiducial model, and the fiducial value of $A_{\mathrm{LSS}}$ is unity. Constraints on $A_{\mathrm{LSS}}$ are equivalent to fractional constraints on $H_0 f \sigma_8$. Note that the transverse velocity from LSS carries the same radial dependence as the SP velocity ($1/r$). Following~\citet{1998PhRvD..57.2117B} and~\citet{2014PhRvD..90f3518H}, we can derive the first iteration of the Newton-Raphson solution of the maximum-likelihood estimator for $A_{\mathrm{LSS}}$ assuming the initial value of $A_{\mathrm{LSS}} = 1$ and replacing the second-derivative of the log-likelihood with its ensemble average, the Fisher information. This gives
\begin{equation}
  \hat{A}_{\mathrm{LSS}} = 1 + \frac{1}{F_{AA}} \left\{ \Delta \mathbf{d}^\dagger \mathbfss{C}_T^{-1}  \mathbfss{C}^u \mathbfss{C}_T^{-1}  \Delta \mathbf{d} - \mathrm{Tr}[\mathbfss{C}_T^{-1}  \mathbfss{C}^u] \right\},
  \label{eq:ALSS}
\end{equation}
where the data vector $\Delta \mathbf{d}$ has potentially had the SAD, SP and correlated LSS proper motions subtracted if we include the $l=1$ mode. This estimator is unbiased, and has variance given by $\sigma_{A_{\mathrm{LSS}}}^2 = F_{AA}^{-1}$, where the Fisher information is
\begin{equation}
  F_{AA} = 2 \mathrm{Tr} \left [\mathbfss{C}_T^{-1}  \mathbfss{C}^u  \mathbfss{C}_T^{-1}  \mathbfss{C}^u  \right].
  \label{eq:FAA}
\end{equation}
The estimator in Equation~\eqref{eq:ALSS} is not the maximum-likelihood estimator, but is optimal in the sense that it has minimum variance, saturating the Cram\'{e}r-Rao bound. In the cosmic-variance limit we have simply $\sigma_{A_{\mathrm{LSS}}} = (4N)^{-1/2}$, which is tiny for our catalogue.

In the LNIN approximation, the Fisher information becomes
\begin{equation}
  F_{AA} = 4 \sum_l \frac{(2l+1)}{2}\mathrm{Tr}\left[(\mathbfss{S}^{E,u}_l + \mathbfss{N}^E)^{-1}\mathbfss{S}^{E,u}_l(\mathbfss{S}^{E,u}_l + \mathbfss{N}^E)^{-1}\mathbfss{S}^{E,u}_l \right],
  \label{eq:FAALNIN}
\end{equation}
which is recognisable from the standard form of the Fisher matrix presented in, e.g.~\citet{1997ApJ...480...22T}. In the LNIN approximation the estimator in Equation~\eqref{eq:ALSS} becomes
\begin{align}
  \hat{A}_{\mathrm{LSS}} &= 1 + \frac{1}{F_{AA}}\left[ \sum_l (2l+1) \right. \nonumber  \\
    & \quad \left. \times \mathrm{Tr}\left\{\left[(\mathbfss{S}^{E,u}_l + \mathbfss{N}^E)^{-1}\hat{\mathbfss{S}}^{E,u}_l - \mathbfss{I} \right](\mathbfss{S}^{E,u}_l + \mathbfss{N}^E)^{-1}\mathbfss{S}^{E,u}_l \right\} \vphantom{\frac12}\right],
  \label{eq:ALSSLNIN}
\end{align}
where $\mathbfss{I}$ is the identity matrix and $\hat{\mathbfss{S}}^{E,u}_l$ is the quadratic estimator for the power spectrum between bin $r_i$ and $r_j$ given by
\begin{equation}
  \left[\hat{\mathbfss{S}}^{E,u}_l\right]_{r_i,r_j} = \frac{1}{2l+1}\sum_m \Delta d^E_{lm}(r_i) \Delta d_{lm}^{E*}(r_j),
\end{equation}
where $d^E_{lm}$ are the E-mode spin-weighted spherical harmonic coefficients of the data. Packaging up the independent elements of the symmetric $N_r \times N_r$ matrix $\mathbfss{S}^{E,u}_l$ into a vector $\mathrm{vecp}(\mathbfss{S}^{E,u}_l)$ of length $N_r(N_r+1)/2$, we can rewrite Equation~\eqref{eq:ALSSLNIN} as
\begin{equation}
  \hat{A}_{\mathrm{LSS}}-1 = \frac{1}{2}\frac{\sum_l \mathrm{vecp}(\mathbfss{S}^{E,u}_l)^\intercal \mathbfss{M}_l^{-1} \mathrm{vecp}(\Delta \mathbfss{S}^{E,u}_l)}{ \sum_l \mathrm{vecp}(\mathbfss{S}^{E,u}_l)^\intercal \mathbfss{M}_l^{-1} \mathrm{vecp}(\mathbfss{S}^{E,u}_l )},
  \label{eq:ALSSLNINalt}
\end{equation}
which follows from Equation~(10) of~\citet{2008PhRvD..77j3013H}. $\mathrm{vecp}(\Delta \mathbfss{S}^{E,u}_l)$ in Equation~\eqref{eq:ALSSLNINalt} is the vectorized debiased quadratic estimator and $\mathbfss{M}_l$ is its covariance matrix (note that for $l\geq2$ we can drop the `$u$' label). The optimal estimator is thus the inverse-variance weighted quadratic estimator for the proper motion power spectrum, cross-correlated with the fixed template spectrum, summed over multipoles and then normalized appropriately, with the factor of $1/2$ present because $A_{\mathrm{LSS}}^2$ is the amplitude of the power spectrum and not $A_{\mathrm{LSS}}$. Note that the $l=1$ moment is contaminated by local terms like SP and SAD, but the estimator is still unbiased when these modes are excluded from both the numerator and denominator of Equation~\eqref{eq:ALSSLNINalt}. Note also that this expression assumes the shape of the template uncorrelated transverse velocity power spectrum is known perfectly, whereas in reality it will depend on the cosmological parameters. We assume that the uncertainty in the cosmology is subdominant to other sources of bias and variance, which is likely to be a very good approximation on linear scales. A more complete treatment would involve simultaneously marginalizing over these in an MCMC chain.

\begin{figure}
  \includegraphics[width=\columnwidth]{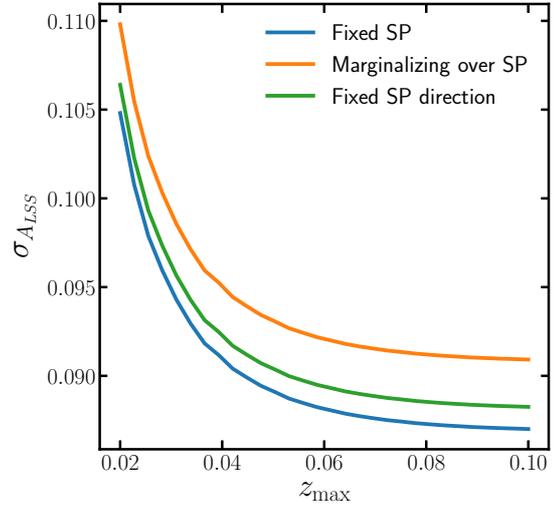}
  \caption{Forecast error on the amplitude of the LSS velocity, using all multipoles (i.e. fixing the SP direction and amplitude; blue, lowest curve), discarding the information from the $l=1$ modes (i.e. marginalizing over the SP direction and amplitude; orange, highest curve), and discarding one third of the information in the $l=1$ modes (i.e. marginalizing over the SP amplitude but keeping its direction fixed; green, middle curve).  The error on $A_{\mathrm{LSS}}$ can be interpreted as the forecast fractional error on the parameter combination $H_0 f \sigma_8$.}
  \label{fig:ALSSSD_zhi}
\end{figure}

In Figure~\ref{fig:ALSSSD_zhi} we plot the forecast error on $A_{\mathrm{LSS}}$ for our assumed survey parameters as a function of $z_{\mathrm{max}}$. The blue (lowest) curve on this plot labelled `Fixed SP' uses all the modes $l\geq1$ in Equation~\eqref{eq:FAALNIN}, while the orange (highest) curve discards the contaminated $l=1$ mode, which is equivalent to marginalizing over the SP proper motion. The green (middle) curve uses only two thirds of the information at $l=1$, which is equivalent to fixing the SP direction and marginalizing over its amplitude. In all cases, the amplitude of LSS transverse velocities should be detectable with $\sigma_{A_{\mathrm{LSS}}} \approx 0.1$, i.e. a significant $10\sigma$ measurement (similar to that claimed in~\citealt{2018arXiv180706670D}) or a 10\% measurement of the parameter combination $H_0f\sigma_8$. That this measurement is more significant than the SP measurement forecast is a testament to the significant amplitude of LSS transverse velocities over this redshift range, and the large amount of information available in the uncontaminated $l\geq2$ modes; only a small ($\sim 5\%$) increase in error is incurred if the $l=1$ mode is discarded. By $z_{\mathrm{max}} = 0.1$ the error has saturated, due to the amplitude of the signal falling away as $1/r$.
%

Unlike the case of the SP constraints in Section~\ref{sec:results}, the constraint on $A_{\mathrm{LSS}}$ is dominated by the instrumental noise (compare Figure~\ref{fig:ALSSSD_zhi} with the cosmic-variance limit of $\sigma_{A_{\mathrm{LSS}}} = (4N)^{-1/2} \approx 10^{-4}$), and thus sensitive to the assumption of point-source proper motion errors for our resolved galaxies. It will therefore be important to thoroughly quantify the impact of proper motion error on these constraints, and we caution the reader against their overinterpretation; our results here show what may be achieved with `perfect' proper motion measurements of galaxies, and thus provide motivation for improving extragalactic astrometry.

\begin{figure}
  \includegraphics[width=\columnwidth]{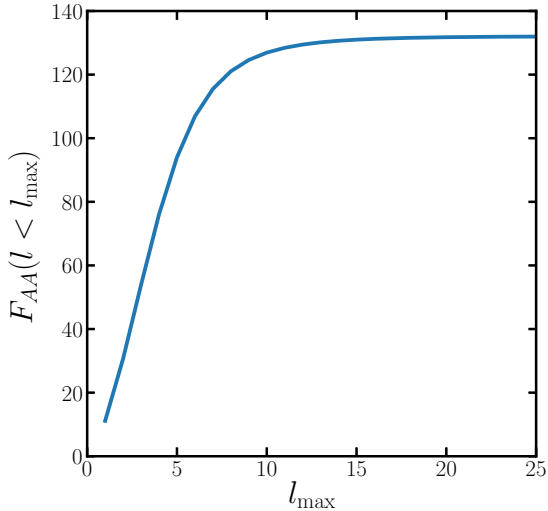}
  \caption{Cumulative Fisher information on the amplitude of the LSS velocity, using only multipoles $l$ less than $l_{\mathrm{max}}$. The error on $A_{\mathrm{LSS}}$ can be interpreted as the forecast fractional error on the parameter combination $H_0 f \sigma_8$.}
  \label{fig:FAA_ells}
\end{figure}

In Figure~\ref{fig:FAA_ells} we plot the cumulative contribution per $l$ to the Fisher information on $A_{\mathrm{LSS}}$. Most of the contribution comes from scales $1 \lesssim l \lesssim 10$ (consistent with Figure~\ref{fig:vperp_ells_xiE}), with the error using only the $l=1$ mode equal to roughly $\sigma_{A_{\mathrm{LSS}}} \approx 0.3$, comparable to the significance forecast for $A_{\mathrm{SP}}$. As found in Section~\eqref{subsec:LSSspec}, these higher $l$-modes can be affected by non-linear corrections to the velocity power spectrum for nearby objects. We defer further investigation into the impact of non-linearity on these forecasts to a future work.

Finally, in Appendix~\ref{app:nvar} we investigate the impact on these forecasts of making different assumptions for the average noise variance; although our choice of using the average inverse variance reproduces the correct answer in the cosmic-variance-free limit for the $A_{\mathrm{SP}}$ estimator, it is unclear whether this choice is correct for $A_{\mathrm{LSS}}$. Using the inverse of the average variance instead of the average inverse variance can increase the forecast uncertainty on $A_{\mathrm{LSS}}$ by a factor of three, which still represents a significant detection. Further investigation using the exact result in Equation~\eqref{eq:FAA} is required, which we again defer to a future work.

\section{Conclusions}
\label{sec:concs}

We have conducted a thorough study into the potential of astrometric surveys to measure the large-scale correlated proper motions of galaxies, focussing on Gaia as a reference survey. Although Gaia is not optimised to measure the positions of extragalactic objects, it will nonetheless produce astrometry and proper motions of roughly $10^6$ galaxies and $10^{5-6}$ quasars~\citep{2012A&A...543A.100R}. With these large catalogues and proper motion errors of $\sim 10^{1-2} \mu\mathrm{as} \, \mathrm{yr}^{-1}$, it is timely to investigate whether this dataset can be used to constrain cosmological parameters.

We have focussed on the large-scale proper motion induced by the Solar System's motion relative to distant galaxies (the secular parallax; SP), the large-scale proper motion caused by the time-varying relativistic aberration from the Solar System's acceleration towards the galactic centre (the secular aberration drift; SAD), and the peculiar transverse velocities of the galaxies induced by large-scale structure (LSS). Extending the previous works of~\citet{2009MNRAS.397.1739D, 2012ApJ...755...58N, 2016A&A...589A..71B, 2018ApJ...864...37D} we have studied the potential to combine extragalactic proper motions with spectroscopic redshifts to measure the Hubble constant through the SP.

We have identified a previously-neglected bias from dipolar transverse velocities correlated with our local motion (similar to the Virgocentric infall and bulk flow phenomena familiar from radial velocity surveys). This bias is really a consequence of attempting to measure only $\boldsymbol{\mu}^{\mathrm{SP}}$, the SP proper motion, rather than $\boldsymbol{\mu}^{\mathrm{SP}} + \boldsymbol{\mu}^{\mathrm{LSS}} = (-\mathbf{v}_{\sun} + \mathbf{v})_{\perp}/r$, which is the \emph{relative} proper motion between the Solar System and the galaxies, and the quantity actually observed. This bias is proportional to the transverse correlation function of~\citet{1988ApJ...332L...7G}, and is especially problematic for nearby objects. As proper motions due to SP and LSS fall away with distance, the cosmic variance and bias from peculiar velocities is potentially very important for this measurement. We found that without proper modelling the bias is important whenever galaxies closer than $50$-$100 \, \mathrm{Mpc}$ are used in the analysis, which limits the significance of the SP velocity to $1$-$2 \sigma$ depending on whether the contaminating SAD proper motion is simultaneously estimated or fixed from a sample of high-redshift quasars. If we add information from the CMB dipole this significance increases to $2$-$4\sigma$, corresponding to a $25$-$50\%$ measurement of $H_0$, with the error dominated by cosmic variance from LSS transverse velocities.

The significance of the $H_0$ measurement could potentially be increased by going to higher redshifts and modelling the bias with perturbation theory or halo models, although the gains appear fairly small. Alternatively, combining with measurements of radial peculiar velocities could help to reduce the cosmic variance. For Gaussian fields at distance $r$, conditioning on the radial velocities of each galaxy would reduce the cosmic variance in $H_0$ by roughly a factor $\rho_1^2(r)$, where $\rho_1(r)$ is the dimensionless correlation coefficient between the part of the radial velocity dipole uncorrelated with our local motion and the part of the transverse velocity dipole uncorrelated with our local motion. Neglecting correlations with our local velocity altogether, we find that $\rho_1^2(r)$ has a broad peak of roughly $0.74$ around $r = 150 \, \mathrm{Mpc}/h$, and is never less than $0.54$ for our galaxies, suggesting that the cosmic variance could be significantly reduced. This simple picture is complicated by the fact that radial velocity surveys typically measure fluctuations in galaxy size or apparent magnitude rather than the radial velocity directly, and the fact that the velocity measurements are noisy, which will degrade the improvements. Nevertheless, the potential gains are large if radial velocities are available.

An accurate and precise measurement of the Hubble constant is vital to break degeneracies with other cosmological parameters such as neutrino mass~\citep{2015PhRvD..92l3535A}, and could shed light on tensions between classical distance ladder and CMB measurements~\citep{2017NatAs...1E.121F}. Since proper motion-based measurements of $H_0$ are robust to the Malmquist and selection biases which complicate radial velocity surveys, further investigation into optimising the measurement is required.

In contrast we find that a significant measurement of the transverse velocities from LSS is possible in our reference Gaia-like survey, forecasting a measurement significance on the amplitude of roughly $10\sigma$, corresponding to a 10\% measurement of the parameter combination $H_0f\sigma_8$, limited by the instrumental noise. This high significance is due to the large amount of information present in the angular structure of the correlation functions of transverse velocities, and the large number of galaxy pairs available to beat down the noise. A measurement of this parameter combination would be extremely useful to cosmology, as the growth rate is sensitive to late-time physics such as dark energy. While the constraints from proper motions are not expected to be competitive with redshift-space distortions in galaxy clustering for some time, an independent measurement of the growth rate serves as a valuable cross-check on the $\Lambda$CDM model and could help pin down some of the systematics in galaxy clustering measurements.

We have also introduced a CMB-style formalism for analysing proper motions on the full sky, which serves as an alternative to the widely-used Vector Spherical Harmonics~\citep{2012A&A...547A..59M, 2018A&A...616A..14G}. This formalism involves expanding the vector-valued proper motion in a basis of orthonormal functions on the sphere which possess the correct rotational properties, the spin-weighted spherical harmonics. Expanding in these functions makes correlating vectors on the sphere straightforward, and gives rise simple analytic formulae which connect correlation functions to angular power spectra. This solves some of the issues raised in~\citet{2018ApJ...864...37D}, who noted that a single correlation function does not contain all the information available in the vector field. We expect this formalism to be of use in any application of proper motion measurements.

Potential improvements to this work include a more realistic noise specification for measuring the proper motion of extended objects such as galaxies, a more realistic treatment of incompleteness and source clustering, and a more thorough treatment of radial peculiar velocities in the distance estimates. In making the large-$N$, isotropic noise (LNIN) approximation we had to assume homogeneous noise variance in each radial bin; the accuracy of this approximation should be tested more rigorously, although we do not expect it to impact the SP forecasts significantly. Firstly these forecasts are limited by cosmic variance, and secondly an estimator with inverse-variance weight containing only instrumental noise is actually fairly close to optimal (compare the green and orange curves in Figures~\ref{fig:SPSD_zhi} and~\ref{fig:ASPSD_zhi}). The detectability of B-mode signatures from gravitational waves or global rotation would also be of interest to study.

This paper provides a necessary step in the development of proper motions from an attractive concept to a feasible and competitive probe of velocities. The potential to test the robustness of the $\Lambda$CDM model with real-time cosmology provides strong motivation for further investigation, which this work anticipates.

\section*{Acknowledgements}

It is my pleasure to thank Anna Lisa Varri for providing useful references, Fergus Cullen and Nigel Hambly for helpful discussions, and Andy Taylor for comments on the manuscript. I would also like to thank the anonymous referee for useful suggestions which improved this paper. I am supported by an STFC Consolidated Grant.




\bibliographystyle{mnras}
\bibliography{references} 




\appendix

\section{Spin-weighted spherical harmonics}
\label{app:formalism}

In this section we detail some formalism surrounding the use of spin-weighted spherical harmonics. These functions were introduced by~\citet{Gelfand},~\citet{doi:10.1063/1.1931221}, and~\citet{doi:10.1063/1.1705135}, and applied to CMB analysis by~\citet{1997PhRvL..78.2054S}. The material here is similar to that presented in~\citet{2014PhRvD..90f3518H}, but we repeat it for completeness.

The transverse velocity field $\mathbf{V}_{\perp}$ is a vector on the sphere, and hence can be decomposed into gradient and curl parts whose components in some coordinate system are
\begin{equation}
  V_{\perp}^a = \nabla^a \Omega + \epsilon^a_{\, \, b} \nabla^b \Psi,
  \label{appeq:vdecomp}
\end{equation}
where $\epsilon^a_{\, \, b}$ is the alternating tensor, $\nabla$ is the covariant derivative on the sphere, and $\Omega$ and $\Psi$ are scalar functions on the sphere. We now introduce the spin raising and lowering operators $\eth$ and $\bar{\eth}$, which act on a spin $s$ quantity ${}_s \eta$ as
\begin{align}
  \eth {}_s \eta &= -(\sin \theta)^s (\partial_{\theta} + i \mathrm{cosec} \, \theta \partial_{\phi}) [(\sin \theta)^{-s} {}_s \eta], \label{appeq:eth1}\\
  \bar{\eth} {}_s \eta &= -(\sin \theta)^{-s} (\partial_{\theta} - i \mathrm{cosec} \, \theta \partial_{\phi}) [(\sin \theta)^{s} {}_s \eta].
  \label{appeq:eth2}
\end{align}
The quantity $\eth {}_s \eta$ has spin $s + 1$ while the quantity $\bar{\eth} {}_s \eta$ has spin $s-1$.

The spin $\pm 1$ spherical harmonics are defined by operating on the standard spherical harmonics as
\begin{align}
  {}_1 Y_{lm} &= \eth \left(\frac{\sqrt{(l-1)!}}{\sqrt{(l+1)!}} Y_{lm}\right), \\
  {}_{-1}Y_{lm} &= -\bar{\eth} \left(\frac{\sqrt{(l-1)!}}{\sqrt{(l+1)!}} Y_{lm}\right).
\end{align}
Note that $Y_{lm} = {}_0 Y_{lm}$. As discussed in Section~\ref{sec:harmonics}, it is desirable to work with quantities with simple transformation properties under a rotation of the coordinate basis. For this reason we work with the complex transverse velocity, found by projection onto the null basis $\hat{\boldsymbol{\theta}} \pm i \hat{\boldsymbol{\phi}}$, given by
\begin{align}
  V_{\theta} + iV_{\phi} &= -\eth(\Omega - i\Psi), \label{appeq:vpm1}\\
  V_\theta -i V_\phi &= -\bar{\eth}(\Omega + i \Psi), \label{appeq:vpm2}
\end{align}
where we used Equations~\eqref{appeq:vdecomp},~\eqref{appeq:eth1} and~\eqref{appeq:eth2}. We now expand the spin 0 potentials $\Omega$ and $\Psi$ in spherical harmonics as
\begin{align}
  \Omega(\hat{\mathbf{n}}) &= \sum_{lm} \sqrt{\frac{(l-1)!}{(l+1)!}} \epsilon_{lm} Y_{lm}(\hat{\mathbf{n}}), \label{appeq:omegalm}\\
  \Psi(\hat{\mathbf{n}}) &= \sum_{lm} \sqrt{\frac{(l-1)!}{(l+1)!}} \beta_{lm} Y_{lm}(\hat{\mathbf{n}}),
\end{align}
where $l \ge 1$ (the $l=0$ mode corresponds to a constant which gives no observable contribution to the transverse velocity). Plugging this into Equations~\eqref{appeq:vpm1} and~\eqref{appeq:vpm2} and using the definition of the spin-weighted spherical harmonics gives
\begin{equation}
  (V_\theta \pm iV_\phi)(\hat{\mathbf{n}}) = \sum_{lm} (\mp \epsilon_{lm} + i\beta_{lm}) {}_{\pm 1} Y_{lm}(\hat{\mathbf{n}}).
\end{equation}
Under a parity transformation, the three-dimensional velocity field transforms as $\mathbf{V}(\hat{\mathbf{n}}) \rightarrow -\mathbf{V}(-\hat{\mathbf{n}})$, such that the transverse components transform as $(V_\theta \pm i V_\phi)(\hat{\mathbf{n}}) \rightarrow -(V_\theta \mp i V_\phi)(-\hat{\mathbf{n}})$. The multipole coefficients thus transform as $\epsilon_{lm} \rightarrow (-1)^l\epsilon_{lm}$ and $\beta_{lm} \rightarrow (-1)^{l+1}\beta_{lm}$, and therefore have electric and magnetic parity respectively. The electric (E-mode) and magnetic (B-mode) coefficients describe gradient and curl contributions to the transverse velocity respectively. Reality of the transverse velocity velocity implies that $\epsilon_{lm}^* = (-1)^m \epsilon_{l-m}$ and $\beta_{lm}^* = (-1)^m \beta_{l-m}$.

The spin-weighted spherical harmonics obey orthogonality and completeness relations over the sphere given by
\begin{align}
  &\int \mathrm{d}\hat{\mathbf{n}} \, {}_s Y_{lm}(\hat{\mathbf{n}}) {}_s Y_{l'm'}^*(\hat{\mathbf{n}}) = \delta_{ll'}\delta_{mm'}, \label{appeq:orthog}\\
  &\sum_{lm} {}_s Y_{lm}^*(\hat{\mathbf{n}}_1) {}_sY_{lm}(\hat{\mathbf{n}}_2) = \delta(\hat{\mathbf{n}}_1 - \hat{\mathbf{n}}_2), \label{appeq:completeness}
\end{align}
where the sums are over $l\ge s$ and $-l \le m \le l$.

The spin-weighted spherical harmonics also obey the relation ${}_s Y^*_{lm} = (-1)^{s+m} {}_{-s}Y_{l-m}$, and satisfy the addition theorem
\begin{equation}
  \sum_m {}_s Y_{lm}^*(\hat{\mathbf{n}}_1)  {}_{s'} Y_{lm}(\hat{\mathbf{n}}_2) = \frac{2l+1}{4\pi}D^l_{ss'}(\gamma_1, \beta_{12}, -\gamma_2),
  \label{appeq:YD}
\end{equation}
where $D^l_{ss'}(\gamma_1, \beta_{12}, -\gamma_2)$ are the Wigner $D$ matrix elements representing a rotation by the Euler angles $\{\gamma_1, \beta_{12}, -\gamma_2\}$ from the $(\hat{\boldsymbol{\theta}},\hat{\boldsymbol{\phi}})$ basis at $\hat{\mathbf{n}}_1$ to that at $\hat{\mathbf{n}}_2$, see for example~\citet{Varsh}.

Under an active rotation of a transverse velocity field by a rotation operator $R$, the scalar potentials $\Omega$ and $\Psi$ must satisfy $[R\Omega](\hat{\mathbf{n}}) = \Omega(R^{-1}\hat{\mathbf{n}})$ and $[R\Psi](\hat{\mathbf{n}}) = \Psi(R^{-1}\hat{\mathbf{n}})$. Use of the matrix elements of the rotation operator in the $(l,m)$-representation (the Wigner $D$ functions) on the spherical harmonics allows us to derive the transformation law for the multipole coefficients
\begin{equation}
  \epsilon_{lm} \rightarrow \epsilon_{lm}' = \sum_{m'} D^l_{mm'}(\alpha, \beta, \gamma)\epsilon_{lm'},
  \label{appeq:rotelm}
\end{equation}
and similarly for $\beta_{lm}$, where $(\alpha, \beta, \gamma)$ are the Euler angles corresponding to $R$. The Wigner $D$ matrix elements are related to the spin-weighted spherical harmonics by
\begin{equation}
  D^l_{-m s}(\phi,\theta,0) = (-1)^m \sqrt{\frac{4\pi}{2l+1}}{}_s Y_{lm}(\theta,\phi),
  \label{appeq:D2SH}
\end{equation}
which can be taken as the definition of the spin-weighted spherical harmonics.

In this paper we are often concerned with dipolar transverse velocities, which correspond to the $l=1$ mode. The explicit expressions for the spin $\pm 1$ spherical harmonics in this case are
\begin{align}
  &{}_1 Y_{10}(\theta, \phi) = \sqrt{\frac{3}{8\pi}} \sin \theta , \label{appeq:1Y0}\\
  &{}_1 Y_{1\pm 1}(\theta, \phi) = -\sqrt{\frac{3}{16\pi}} (1 \mp \cos \theta) e^{\pm i\phi}. \label{appeq:1Y1}
\end{align}

We close this section by connecting the spin-weighted spherical harmonics to the widely used vector spherical harmonics (VSHs). These are defined from the standard spherical harmonics as~\citep{2012A&A...547A..59M}
\begin{align}
  \mathbf{S}_{lm} &= \frac{1}{\sqrt{l(l+1)}} \nabla Y_{lm}, \\
  \mathbf{T}_{lm} &= -\hat{\mathbf{n}} \times \mathbf{S}_{lm},
\end{align}
where formally $\nabla$ is now the three-dimensional gradient operator projected orthogonal to the unit sphere, and $\mathbf{S}_{lm}$ and $\mathbf{T}_{lm}$ are three-dimensional vectors orthogonal to $\hat{\mathbf{n}}$. Using the definition of the spin $\pm 1$ spherical harmonics, it is straightforward to show that
\begin{align}
  (\hat{\boldsymbol{\theta}} \pm i \hat{\boldsymbol{\phi}})\cdot \mathbf{S}_{lm} &= \mp {}_{\pm 1} Y_{lm}, \label{appeq:Slm}\\
  (\hat{\boldsymbol{\theta}} \pm i \hat{\boldsymbol{\phi}})\cdot \mathbf{T}_{lm} &= i {}_{\pm 1} Y_{lm}. \label{appeq:Tlm}
\end{align}
The spin $\pm 1$ spherical harmonics are thus projections of the VSHs onto a null basis. The expansion of a vector field on the sphere in VSHs is
\begin{equation}
  \mathbf{V} = \sum_{lm} (s_{lm} \mathbf{S}_{lm} + t_{lm} \mathbf{T}_{lm}).
\end{equation}
Projecting onto the null basis and using Equations~\eqref{appeq:Slm} and~\eqref{appeq:Tlm} allows us to identify the coefficients of the VSH expansion with those of the spin-weighted spherical harmonic expansion as $(s_{lm}, t_{lm}) = (\epsilon_{lm}, \beta_{lm})$. We advocate the use of spin-weighted spherical harmonics over VSHs, as they are much easier to work with when correlating transverse vector fields over the full sky.
  
\section{Relationship of $\xi_\pm$ to the correlation functions of Darling \& Truebenbach 2018}
\label{app:darling}

In this section we compare our $\xi_\pm$ correlation functions to those recently introduced in~\citet{2018ApJ...864...37D} (hereafter DT18). Their Equation~(1) defines a correlation function
\begin{equation}
  \xi_{v_\perp}(\mathbf{x}_1, \mathbf{x}_2) = \langle [\mathbf{V}_\perp(\mathbf{x}_1) \cdot \hat{\mathbf{x}}] [\mathbf{V}_\perp(\mathbf{x}_2) \cdot \hat{\mathbf{x}}] \rangle,
\end{equation}
where $\mathbf{x} = \mathbf{x}_1 - \mathbf{x}_2$. To relate this correlation function to $\xi_\pm$, we first resolve $\hat{\mathbf{x}}$ at the point $\mathbf{x}_1$ into a part parallel to $\mathbf{x}_1$ and a part orthogonal to $\mathbf{x}_1$; only the orthogonal part survives contraction with $\mathbf{V}_\perp(\mathbf{x}_1)$. We then rotate to the plane containing the origin and the vectors $\mathbf{x}_1$ and $\mathbf{x}_2$. $\mathbf{V}_\perp(\mathbf{x}_1) \cdot  \hat{\mathbf{x}}$ is then the component of $\mathbf{V}_\perp(\mathbf{x}_1)$ in this plane, and may be related to the $\theta$-component in the spherical coordinate basis aligned with the plane by
\begin{equation}
  \mathbf{V}_\perp(\mathbf{x}_1) \cdot \hat{\mathbf{x}} = \frac{\lvert \mathbf{x}_2 - (\mathbf{x}_2 \cdot \hat{\mathbf{x}}_1)\hat{\mathbf{x}}_1 \rvert}{\lvert \mathbf{x}_2 - \mathbf{x}_1\rvert} \bar{V}_\theta(\mathbf{x}_1),
  \label{appeq:vperpdotx}
\end{equation}
where an overbar indicates $\mathbf{V}_\perp$ in the coordinate basis rotated into the plane, as per the discussion in Section~\eqref{sec:harmonics}. The prefactor in Equation~\eqref{appeq:vperpdotx} may be written in terms of the radial distances of the point $r_1$ and $r_2$ and the angle between them $\beta_{12}$ as
\begin{equation}
  \frac{\lvert \mathbf{x}_2 - (\mathbf{x}_2 \cdot \hat{\mathbf{x}}_1)\hat{\mathbf{x}}_1 \rvert}{\lvert \mathbf{x}_2 - \mathbf{x}_1\rvert} = \frac{r_1 r_2 \sin^2 \beta_{12}}{r_1^2 + r_2^2 - 2r_1r_2\cos \beta_{12}} \equiv f(\beta_{12},r_1,r_2).
\end{equation}
Similar relations apply at $\mathbf{x}_2$. Using $V_{\pm} = V_\theta \pm i V_\phi$ to relate $\bar{V}_\theta$ to $V_{\pm}$, and the definition of $\xi_\pm$, we find that
\begin{equation}
  \xi_{v_\perp}(\mathbf{x}_1, \mathbf{x}_2) = f(\beta_{12},r_1,r_2)\frac{1}{2} \left[\xi_+(\beta_{12},r_1,r_2) + \xi_-(\beta_{12},r_1,r_2)\right].
  \label{appeq:DT1}
\end{equation}
In other words, the correlation function $\xi_{v_\perp}$ defined in~DT18 is proportional to a linear combination of our $\xi_\pm$ correlation functions. Note that Equation~\eqref{appeq:DT1} is independent of the orientation of the coordinate basis (as required) but only contains half of the total information available.

DT18 also define a correlation function
\begin{equation}
  \xi'_{v_\perp}(\mathbf{x}_1,\mathbf{x}_2) = \langle \mathbf{V}_\perp(\mathbf{x}_1) \cdot \mathbf{V}_\perp(\mathbf{x}_2) \rangle,
\end{equation}
where the vectors here are assumed to be embedded in three-dimensional space. Using similar geometric arguments to above, we find that
\begin{align}
  \xi'_{v_\perp}(\mathbf{x}_1,\mathbf{x}_2) &=  \left(\frac{1 + \cos \beta_{12}}{2}\right)\xi_+(\beta_{12},r_1,r_2) \nonumber \\
  & \, + \left(\frac{-1 + \cos \beta_{12}}{2}\right)\xi_-(\beta_{12},r_1,r_2),
  \label{appeq:DT2}
\end{align}
i.e. again a linear combination of the $\xi_\pm $ correlation functions. Once more, individually this correlation function only contains half the information available in the vector field; use of the $\xi_\pm $ correlation functions ensures no information is lost. In addition the $\xi_\pm $ are related straightforwardly to the power spectra of the E and B modes as shown in Section~\ref{sec:harmonics}, and hence can be easily related to the VSH description.

Finally, it may be shown that Equations~\eqref{appeq:DT1} and~\eqref{appeq:DT2} when applied to the linear velocity field of large-scale structure agree with the expressions in~DT18 upon use of our Equation~\eqref{eq:zetaE}, the explicit forms of the Wigner $d$ elements given in~\citet{Varsh}, and the useful expression
\begin{equation}
  \sum_{l=0}^\infty (2l+1)j_l(kr_1) j_l(k r_2) P_l^{(n)}(\cos \beta) = \frac{(k r_1)^n (k r_2)^n}{(kx)^n} j_n(kx),
  \label{appeq:greateq}
\end{equation}
where $ P_l^{(n)}$ is the $n$-th derivative of the Legendre polynomial with $n \ge 0$, $j_l$ is a spherical Bessel function, and $x \equiv (r_1^2 + r_2^2 - 2r_1r_2\cos \beta)^{1/2}$. The proof of Equation~\eqref{appeq:greateq} follows by induction.

\section{Alternative choices for the average noise variance}
\label{app:nvar}

When inverting the covariance matrix $(\mathbfss{N} + \mathbfss{C})$, we had to assume that the noise variance was constant in each radial bin. As discussed in Section~\ref{subsec:LNIN} this is not correct, but using the average value of the inverse noise variance $\langle \sigma_\mu^{-2} \rangle$ reproduces the exact result in the limit of zero cosmic variance. In reality the noise variance varies due to the distribution of magnitude and colour in each bin (c.f. Figure~\ref{fig:gums_gals_pmerrdist}), and a proper treatment would involve marginalizing over the distributions of these latent variables. This must be done numerically since the distributions are all non-Gaussian and the dependencies are non-linear. In this section we take a simpler approach and just study the impact of different choices for the average noise variance in the noise power spectrum on the forecasts. Specifically, we investigate the impact of using the inverse of the average noise variance, $\langle \sigma_\mu^2 \rangle^{-1}$ instead of  $\langle \sigma_\mu^{-2} \rangle$.

When using $\langle \sigma_\mu^2 \rangle^{-1}$ and all the galaxies with $r \geq 20 \, \mathrm{Mpc}/h$ and $z \leq 0.1$, we find that the forecast error on a single component of the SP velocity increases by roughly 20\% (marginalized SAD) and 50\% (fixed SAD), with similar changes to the amplitude parameter $A_{\mathrm{SP}}$. These mild changes are due to the variance in these estimators being dominated by cosmic variance from LSSt transverse velocities rather than the instrumental noise (c.f. Figure~\ref{fig:SPSD_zhi}), so the results of Section~\ref{sec:results} should be robust to assumptions about the average noise variance.

%
In contrast, the forecasts on $A_{\mathrm{LSS}}$ are dominated by instrumental noise, and are quite sensitive to assumptions about the average noise variance. Using the inverse-average instead of the average-inverse, we find the size of the forecast error of $A_{\mathrm{LSS}}$ is larger by roughly a factor of three, suggesting that $\sigma_{A_{\mathrm{LSS}}} \approx 0.3$ for this choice, still a significant measurement. This suggests that more work will be required to investigate the impact of inhomogeneous noise on the $A_{\mathrm{LSS}}$ estimator variance, which we defer to a future work.


\bsp	
\label{lastpage}
\end{document}